\documentclass[11pt,onecolumn,oneside]{osajnl}
\usepackage{amssymb}
\usepackage{amsmath}
\usepackage{hyperref}

\journal{jocn} 

\title{Security Analysis in Multicasting over Shadowed Rician and $\alpha$-$\mu$ Fading Channels: A Dual-hop Hybrid Satellite Terrestrial Relaying Network}

\author[1]{A. S. Sumona}
\author[2]{Milton Kumar Kundu}
\author[3]{A. S. M. Badrudduza}

\affil[1,3]{Department of Electronics \& Telecommunication Engineering, Rajshahi University of Engineering \& Technology (RUET), Rajshahi-6204, Bangladesh}
\affil[2]{Department of Electrical \& Computer Engineering, RUET}

\begin{abstract}
In this era of 5G technology, the ever-increasing demands for high data rates lead researchers to develop hybrid satellite-terrestrial (HST) networks as a substitution to the conventional cellular terrestrial systems. Since an HST network suffers from a masking effect which can be mitigated by adopting the terrestrial relaying strategy, in this work, we focus on wireless multicasting through an HST relaying network (HSTRN) in which a satellite sends messages to multiple terrestrial nodes via multiple relays under the wiretapping efforts of multiple eavesdroppers. Our concern is to protect the multicast messages from being eavesdropped taking advantage of the well-known opportunistic relaying technique. We consider the satellite links follow Shadowed Rician fading whereas the terrestrial links undergo $\alpha-\mu$ fading. The secrecy performance of the proposed HSTRN model is accomplished by deriving expressions for the probability of nonzero secrecy multicast capacity, ergodic secrecy multicast capacity, and secure outage probability for multicasting in closed-form. Capitalizing on the derived expressions, we analyze how a perfect secrecy level can be preserved in spite of harsh channel conditions and also present a secrecy trade-off in terms of the number of relays, multicast users, and multiple eavesdroppers. Finally, the numerical analyses are corroborated via Monte-Carlo simulations.

\quad

Keywords: $\alpha-\mu$ fading, secure multicasting, satellite terrestrial network, shadowed Rician, secure outage probability.
\end{abstract}

\begin{document}

\maketitle

\section{Introduction}

\subsection{Background}

Satellite communication (SatCom) has gained enormous popularity because of its' capability for uninterrupted high-speed data transfer covering a very large area. This means of the communication system can be employed in navigation, disaster management, and those places in the world where the application of wired or wireless terrestrial communication network is uneconomical. As the total number of users around the world is growing every day, the licensed frequency spectrum is becoming inadequate. To reduce this problem, researchers have proposed a hybrid framework of satellite-terrestrial (ST) network where both satellite and terrestrial systems share the frequency spectrum to maximize its' utilization \cite{an2016secure}. Wireless multicasting can also contribute to the maximum utilization of spectrum sharing as it is used to send the same data at the same time to a group of selected users without sending replicas of the same signal to all individually. So multicasting through hybrid ST communication link can be a breakthrough in wireless communication technology.

\subsection{Related works}

One of the main concerns in SatCom is the shadowing of the ST link which is responsible for the masking effect and results in a blockage of the Line-of-Sight (LOS) communication path. The effect of shadowing on wireless communication network was analyzed in \cite{bandjur2013second, simmons2018double, aljuneidi2016approximating, gonzalez2015mimo, 8107558} where the authors in \cite{bandjur2013second} defined shadowing as the result of large obstacles and large deviations present between transmitter and receiver, and analyzed its effect on the received signal with microdiversity and macro diversity reception over composite shadowed Rician (SR)-Gamma fading channel. 
The Double SR model was explored in \cite{simmons2018double} for two formats of fluctuation and shadowing model. A further effective model named Loo's Model was analyzed in \cite{aljuneidi2016approximating} to analyze its Line-of-sight (LoS) shadowing performance. Shadowing can be not only a foe but also a friend for the wireless system which was proved in \cite{gonzalez2015mimo}. Here the authors proved that the capacity of a distributed multiple-input multiple-output(D-MIMO) Rayleigh-Rician channel can be improved with the increasing amount of shadowing. Generalization of composite multipath shadowing model can also be obtained from the product of two fading variables \cite{8107558}.
The effect of both shadowing and correlation was analyzed over a D-MIMO Rician/Gamma fading channel in \cite{li2017performance}, where the authors numerically analyzed how they can affect the achievable sum rate (ASR), symbol error ratio (SER), and outage probability (OP) of the system. The effect of correlation and shadowing can be overcome by diversity techniques, such as maximal ratio combining (MRC) and selection combining (SC). Authors in \cite{sekulovic2012performance} exhibited how these techniques can compensate for the deleterious effect of correlation.

Due to the inherent broadcast nature of wireless links, secrecy analysis over wireless networks is always a key part of the research. The conventional secrecy analyses in the upper layer were conducted utilizing cryptographic codes which started diminishing its attraction with the rapid arrivals of powerful eavesdroppers. Hence, secrecy analysis at the physical layer taking advantage of the random characteristics of the propagation link was first introduced in Wyner's Model \cite{wyner1975wire} which is now well-known as the information-theoretic security model. The average secrecy capacity (ASC) of $\alpha-\mu$ fading environment was analyzed in \cite{7856980} and the secure outage probability (SOP) and the probability of non-zero secrecy capacity (PNZSC) of double SR fading channel were determined in \cite{ai2019secrecy} considering this model. The detrimental effect of correlation over the secrecy performance was also studied in \cite{le2018performance} for the Rician fading channel where the author determined the closed-form expressions of ergodic secrecy capacity (ESC) and SOP. Researchers have proved that physical layer security (PLS) clearly outperforms the standard encryption method \cite{tsouri2012threshold} and performance of PLS (such as secrecy throughput) for different conditions has also been studied \cite{pattanayak2020physical}. Authors in \cite{li2011ergodic,wang2015jamming} analyzed the information security over Rician fading channel where in \cite{li2011ergodic}, they have found the optimal input covariance for the highest ergodic secrecy rate (ESR) and proposed Newton-type method and piyavskii's algorithm as local and global maximizer, respectively. The geometric locations of the receiver and eavesdropper are also an important issue and it was highlighted in \cite{wang2015jamming} for massive MIMO systems via AN-aided jamming. Security of satellite wiretap SR channel was also analyzed in \cite{guo2018secrecy} by evaluating the expressions of PNZSC, SOP, and ASC. The probability of strictly positive secrecy capacity (SPSC) is evaluated in \cite{liu2012probability} for the composite fading scenarios of Rayleigh/Rayleigh, Rician/Rayleigh, Rayleigh/Rician, and Rician/Rician. Considering two homogeneous Poisson processes (HPPPs), analytical expressions of connection outage probability (COP), PNZSC, and ESC were also determined in \cite{8477185} for $\alpha-\mu$ fading channel.
The beamforming approach can increase the PLS over the wireless channel which in turn reduces the SOP \cite{liu2016location} and also extends the network range without compromising channel security \cite{yan2015location}.
Secrecy performance can also be enhanced by employing the MIMO system over generalized $\alpha-\mu$ channel as the diversity and antenna gain increases the PNZSC and ASC and decreases the SOP \cite{moualeu2019transmit}.

Later, researchers started working on cooperative networks to increase the data security and reliability of the wireless communication system. The advantages of the cooperative network were analyzed in  \cite{guo2017outage} in terms of OP and asymptotic OP for both Amplify-and-forward (AF) and Decode-and-forward (DF) protocol. A similar analysis was done in \cite{magableh2014capacity,aldalgamouni2017outage} over $\alpha$-$\mu$ fading channel where in \cite{magableh2014capacity}, end to end capacity and outage capacity of the system was derived and \cite{aldalgamouni2017outage} represents the analytical approach for end-to-end OP and moment generating function (MGF).
The effect of correlation degrades the overall performance of cooperative networks \cite{han2016outage} which can be compensated by employing the opportunistic relaying technique.
The security over the cooperative network was analyzed in \cite{zhang2018secure,ahmed2018secrecy,8354927} where in \cite{zhang2018secure}, authors considered multi-pair massive MIMO relaying networks over Rician fading channels while determining its security. An AN aided security technique was proposed in \cite{ahmed2018secrecy} where the quality of service was explored in a secure cooperative network. Later on,  authors in \cite{8354927} provides a reliability and security analysis over cascaded $\alpha$-$\mu$ fading channels.

As satellite has numerous advantages over other means of communication systems, researchers have also studied the performance of the ST network employing cooperative relays in \cite{lin2020integrated, huang2019performance, bu2016performance, liu2019performance, zhao2017ergodic}.
The security and reliability of a 5G-satellite network was studied in \cite{lin2020integrated} where the authors outlined the future trends of the 5G-satellite communication network.
Performance of a DF relay assisted ST network was analyzed in \cite{huang2019performance} considering the scenario where the satellite-relay and relay-user link undergoes different fading conditions.
A similar analysis was done for fixed gain AF relay in \cite{bu2016performance} and derived the systems EC and average symbol error rate (ASER).
A moving unmanned aerial vehicle (UAV) with multiple antennas was engaged as the cooperative relay in \cite{liu2019performance} to evaluate the outage probability (OP) in an ST network.
The advantages of multiple relays in an ST communication system were examined in \cite{zhao2017ergodic} in terms of ergodic capacity (EC).
The secrecy performance of the ST network was studied in \cite{guo2016secure,miridakis2014dual} considering SR fading distribution.
An energy efficient secure ST cognitive network was considered in \cite{lin2020secure}. The authors here used Rate-splitting multiple access technique to achieve a superior result. 
The authors in \cite{huang2021uplink} developed a space division multiple access scheme for a mixed RF/FSO based ST network with a view to maximize the ergodic sum rate assuming  imperfect angular information of all the user equipment.
A comprehensive secrecy performance analysis of a multiuser-based Hybrid ST Relay Network (HSTRN) has been explained in \cite{bankey2018physical} by adopting SR fading for satellite link and Nakagami-m fading for terrestrial links. For security measurement, authors derive an accurate and asymptotic SOP expression.
In addition, the impact of hardware impairments and co-channel interference (CCI) on the security of satellite relay communication network in SR channel has been presented in \cite{kefeng2017performance}.


\subsection{Motivation}
Researchers in \cite{bandjur2013second,aljuneidi2016approximating,simmons2018double,gonzalez2015mimo,8107558} only described the effect of shadowing on the performance of the wireless communication system. The detrimental effects of correlation and shadowing were analyzed in \cite{li2017performance, sekulovic2012performance}. The PLS of wireless system was evaluated in \cite{7856980,ai2019secrecy,le2018performance,tsouri2012threshold,pattanayak2020physical,li2011ergodic,wang2015jamming,guo2018secrecy,liu2012probability,8477185,liu2016location,yan2015location,moualeu2019transmit} considering only single receiver and eavesdropper. The properties and advantages of a cooperative network with a single relay were illustrated in \cite{guo2017outage,magableh2014capacity,aldalgamouni2017outage}  and their security issue was investigated in \cite{zhang2018secure,ahmed2018secrecy,8354927}. The application of satellite communication was introduced in \cite{huang2019performance,bu2016performance,liu2019performance,zhao2017ergodic} with a cooperative wireless network, and in \cite{guo2016secure,miridakis2014dual,bankey2018physical}, the authors analyzed the security aspect of such network. So it can be seen that no literature mentioned above has considered the possibility of multicasting or the presence of multiple eavesdroppers in the system. Although an opportunistic relaying scheme was applied in \cite{han2016outage}, the authors did not consider the multicasting scenario or shadowing in the environment. According to the best of the author's knowledge, no research has yet been done to measure the secrecy capacity of a wireless network employing the best relay selection scheme from multiple relays with shadowing at one side and multiple receivers and eavesdroppers on the other side considering the generalized fading condition in the relay to receivers/eavesdroppers link. This research gap fulfilled in this paper by considering a dual-hop satellite multicast communication network with multiple relays and eavesdroppers considering SR fading in the first hop and $\alpha-\mu$ fading channel in the second hop. The reason behind the performance evaluation of a generalized fading channel considered in the second hop is that it can be used to analyze some well-known classical distribution models such as one-sided Gaussian, Rayleigh, Weibull, and Nakagami-$m$ fading models \cite{lei2017secrecy}.


\subsection{Contribution}
The key contributions of this work are summarized as follows:
\begin{enumerate}
    \item A cooperative dual-hop model with multiple relays and receivers in the presence of multiple eavesdroppers is considered in this proposed work which is a novel approach in satellite-terrestrial networking system since no previous works considered multicasting under the multiple eavesdroppers' wiretapping attempts over such networks. 
    
    \item At first, we derive dual-hop PDFs considering the best relay selection criteria by using the probability density functions (PDFs) of each individual hop. Capitalizing on those PDFs, we derive PDFs for multicast and eavesdropper channels using order statistics which are also novel with respect to the previous studies. 

    \item To evaluate the security aspects of the proposed model, the closed-form expressions of secure outage probability for multicasting (SOPM), ergodic secrecy multicast capacity (ESMC), and the probability of non-zero secrecy multicast capacity (PNSMC) have been derived and verified through Monte-Carlo simulations.

    \item Although some researches on the secrecy analysis of satellite communications are reported in the existing works, (e.g. \cite{huang2017secrecy, cao2018relay, bankey2017secrecy, bankey2019physical}), it is noteworthy that our proposed model clearly establishes a superiority over those works by replicating most of those existing works as our special cases.
\end{enumerate}


\subsection{Arrangement of the paper}
The paper has been rearranged as follows. System model and problem formulation are demonstrated in section \ref{sec2}. The analytical expressions of SOPM, ESMC, PNSMC are derived in section \ref{sec3}, \ref{sec4} and \ref{sec5}, respectively. Numerical analysis of the derived expressions is elaborately described in section \ref{sec6}. Lastly, this research is summarized in section \ref{sec7}.


\section{System Model and Problem Formulation}
\label{sec2}

\begin{figure}[!ht]
\vspace{0mm}
    \centerline{\includegraphics[width=0.455\textwidth]{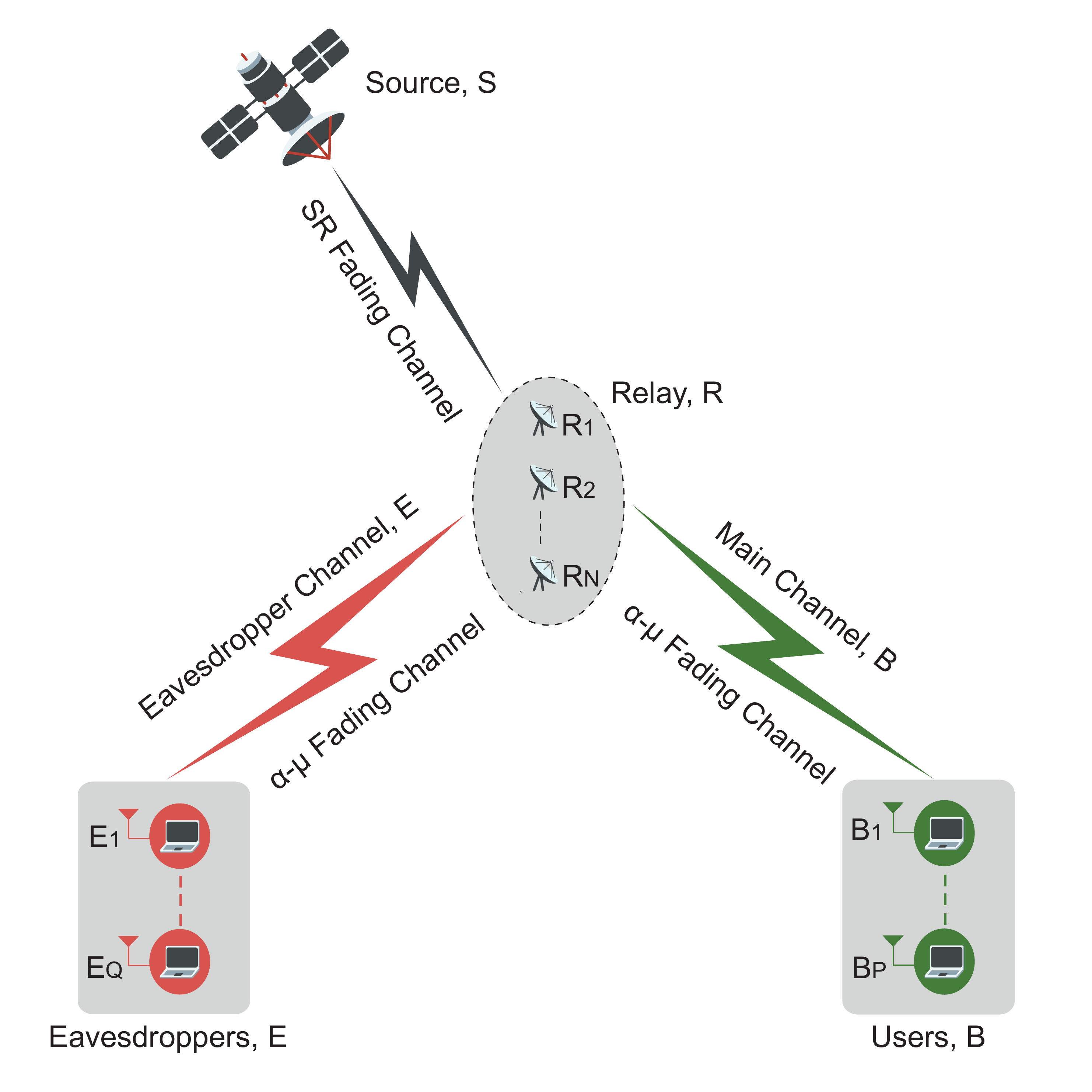}}
        \vspace{0mm}
    \caption{Relay selection in multicasting over a dual-hop HSTRN system in the presence of multiple eavesdroppers.}
    \vspace{0mm}
    \label{Fig.eps}
\end{figure}
A real-life scenario is considered in Fig. \ref{Fig.eps} where a satellite source, $S$ intends to send some private data to a bunch of selected users, $B$ which are resided at a distant location from $S$. Due to this long-distance between $S$ and $B$, there exists no direct communication path between them. Therefore, the relay is employed which will collect the information from $S$, process it, and transmit that same information again to $B$. This model can be practically applied to video conferencing or distance learning \cite{badrudduza2020enhancing} through device-to-device (D2D) and mmWave communication networks in the widely known 5G system and beyond \cite{kong2018secrecy}. As multiple eavesdroppers, $E$ are present in the system, they will try to overhear that private data sent from the relay. Thus security of the transmitted data will be jeopardized. The enhancement of the information security can be achieved by engaging multiple relays ($N$) instead of a single relay and choosing the best one among all the $N$ relays commonly known as the best relay selection scheme.

So, there are actually two hops of communication in this proposed model. In the first hop, the data is sent from $S$ to the relays via the $\mathcal{S}-\mathcal{N}$ link which is modeled to experience the SR fading. The actual cause of considering the SR fading channel is to evaluate the effect of shadowing on the secrecy performance in satellite communication. Then the best relay is chosen among the given $N$ relays and the data is sent from this best relay to the users, $B$ through best relay-user, $\mathcal{N}-\mathcal{P}$  link which is designed as the $\alpha-\mu$ fading channel because this simple yet flexible model can be used to operate in different environmental condition as required in 5G communications \cite{freitas2017complex}. The eavesdroppers, $E$ are wiretapping the transmitted data via $\mathcal{N}-\mathcal{Q}$ link. All the nodes are designed with a single antenna.

Let, the environmental or channel condition of the $\mathcal{S}-\mathcal{N}$ link is represented by the coefficient, $u_{sr}\in \mathcal{C}^{1\times1}$ where ($r=1,2,3,\dots,N$). Similarly, the channel condition for $\mathcal{N}-\mathcal{P}$ and $\mathcal{N}-\mathcal{Q}$ link can be given by the coefficients $v_{rb}\in \mathcal{C}^{1\times1}$ ($b=1,2,3,\dots,P$) and $w_{re}\in \mathcal{C}^{1\times1}$ ($e=1,2,3,\dots,Q$). So the relay will receive a signal of the shape given as following:
\begin{align}
d_{sr}=u_{sr}z+n_{r}.
\end{align}
Here, $z\sim\widetilde{\mathcal{N}}(0,N_s)$ is the transmitted signal, where $N_s$ is the transmit power. The noise term at $r$th relay is denoted by $n_{r}\sim\widetilde{\mathcal{N}}(0,N_r)$, where $N_{r}$ is the noise power. In the second hop, the signal from the best relay will be received by $P$ and $Q$. So, the signal received at $b$th users and $e$th eavesdroppers can be expressed as
\begin{align}
d_{rb}&=v_{rb}d_{r}+x_{b}=v_{rb}(u_{r}z+n_{r})+x_{b}=p_{rb}z+g_{b},
\\
d_{re}&=w_{re}d_{r}+y_{e}=w_{re}(u_{r}z+n_{r})+y_{e}=p_{re}z+h_{e},
\end{align}
where $g_{b} \triangleq v_{rb}n_{r}+x_{b}$, $h_{e} \triangleq w_{re}n_{r}+y_{e}$, $x_{b}\sim\widetilde{\mathcal{N}}(0,N_{b})$ and $y_{e}\sim\widetilde{\mathcal{N}}(0,N_{e})$ describe Gaussian noises, $N_{b}$ and $N_{e}$ represents the noise power at the $b$th users and $e$th eavesdroppers, respectively. Now the instantaneous SNRs of the $\mathcal{S}-\mathcal{N}, \mathcal{N}-\mathcal{P}$ and $\mathcal{N}-\mathcal{Q}$ links can be written as $\xi_{sr}={\frac{N_s}{N_{r}}}\|u_{sr}\|^{2}$, $\xi_{rb}={\frac{T_p}{N_{b}}}\|v_{rb}\|^{2}$ and $\xi_{re}={\frac{T_p}{N_{e}}}\|w_{re}\|^{2}$, respectively, where the transmit power from $r$th relay is denoted as $T_p$.


\subsection{PDFs of SR and $\alpha$-$\mu$ fading channels}
Since the first hop of this system is considered to experience the SR fading channel, the PDF of $\xi_{{s}{r}}$ for the $\mathcal{S}-\mathcal{N}$ link can be given as \cite{guo2016secure}
\begin{align}
\label{4}
f_{\xi_{{s}{r}}}(\xi)=\lambda_{s}e^{{-{\chi_{s}}}{\xi}}{_{1}F_{1}}\left(m_{s}; 1; {\frac{\gamma_{s}}{\varrho_{s}}}\xi\right),
\end{align}
where $\chi_{s}=\frac{1}{{2P_{s}}{\varrho_{s}}}$, $\lambda_{s}={(\frac{2P_{s}{m_{s}}}{2P_{s}{m_{s}}+\xi_{s}})^{m_{s}}}{\chi_{s}}$, $\gamma_{s}=\frac{\xi_{s}}{{2P_{s}}({2P_{s}{m_{s}}+\xi_{s}})}$, $2P_{s}$ represents the average power of the multipath component, $m_{s}\geq{0}$ is the fading severity parameter ranging from 0 to 1 and $\xi_{s}$ is the average power of the line-of-sight (LOS) component. $_{1}F_{1}{(.;.;.)}$ presents the confluent hyper-geometric function and the average SNR is represented by $\varrho_{s}$. The amount on shadowing in the SR Fading channel depends on the values of $P_s, m_s$ and $\xi_{s}$. These parameters can be easily calculated from Loo's model using the following equations given by \cite[eqs.~10,11,]{1198102}
\begin{align}
\label{5}
    &\mu =\frac{1}{2}\Biggl[\ln\Big(\frac{  \xi _s}{m_s}\Big)+\psi  (m_s)\Biggl],
    \\
    \label{6}
   & d_0 =\frac{\psi ^{'}(m_s)}{4},
\end{align}
where $\mu$ and $d_0$ are the Loo's parameter. $\psi(.)$ and $\psi ^{'}(.)$ represents Psi function \cite[eq.~8.360.1,]{GR:07:Book} and its first derivative, respectively. For a given value of $d_0$, $m_s$ can be easily calculated from \eqref{6}. The numerical value of $\xi_{s}$ can be easily found from \eqref{5} using the value of $m_s$. Four different cases of shadowing for some specific values of these parameters are given in Table \ref{table1} where case 1 represents the heavy shadowing and case 4 represents very light shadowing condition.
\begin{table}[!ht]
\centering
\caption{Parameter values ($P_s, m_s$ and $\xi_{s}$) calculated from Loo's parameters using \eqref{5} and \eqref{6}.}
\label{table1}
\begin{tabular}{c|c|c|c|c|c}
\hline
\multirow{2}{*}{\begin{tabular}[c]{@{}c@{}}Shadowing 
\\ 
Cases\end{tabular}} & \multicolumn{2}{c|}{Loo's Model} & \multicolumn{3}{c}{Proposed Model} 
\\ 
\hline
 & $\mu$        & $\sqrt{d_0}$      & $P_s$     & $m_s$     & $\xi_s$     
\\ 
\hline
\hline
Case-1   & $-3.914$       & $0.806$             & $0.063$     & $0.739$     & $0.0009$      
\\ 
\hline
Case-2  & $-0.690$       & $0.230$             & $0.251$     & $5.21$      & $0.278$       
\\ 
\hline
Case-3   & $-0.115$       & $0.161$             & $0.126$     & $10.1$      & $0.835$       
\\ 
\hline
Case-4  & $0.115$        & $0.115$             & $0.158$     & $19.4$      & $1.29$       
\\ 
\hline
\end{tabular}
\end{table}

Now, applying \cite[eq.~9.14.1,]{GR:07:Book} in (\ref{4}), we get
\begin{align}
\label{7}
f_{\xi_{{s}{r}}}(\xi)= \sum _{\theta _1=0}^{\infty }\psi_{s}\xi ^{\theta _1} e^{-\chi _s  \xi},
\end{align}
where $\psi_{s}=\lambda _s \frac{\gamma _s^{\theta _1} \left(m_s\right)_{\theta _1}}{\left(\theta _1!\right){}^2 \varrho _s^{\theta _1}}$ and $({m_{s}})_{\theta_{1}}$ is the Pochhammer symbol where $({m_{s}})_{\theta_{1}}=\frac{\Gamma({m_{s}}+\theta_{1})}{\Gamma({m_{s}})}$.

In the second hop of the proposed system, both the $\mathcal{N}-\mathcal{P}$ link and $\mathcal{N}-\mathcal{Q}$ link follows the $\alpha-\mu$ fading distribution. So the PDF of $\xi_{{r}{b}}$ and $\xi_{{r}{e}}$ for $\mathcal{N}-\mathcal{P}$ link and $\mathcal{N}-\mathcal{Q}$ can be respectively given by \cite{7856980}
\begin{align}
\label{8}
f_{\xi_{{r}{i}}}(\xi)={z_{1}}e^{{-\psi_{1}}{\xi^{\frac{\alpha_{i}}{2}}}}{\xi^{z_{2}}},
\end{align}
where $z_{1}={\frac{{\alpha_{i}}{\mu_{i}}^{{\mu}_{i}}}{2\Gamma{{({\mu}_{i})}}}}{\varrho_{i}}^{-\frac{{\alpha_{i}}{\mu_{i}}}{2}}$,  $\psi_{1}={\mu_{i}}{{\varrho_{i}}^{-\frac{\alpha_{i}}{2}}}$, $z_{2}={\frac{{\alpha_{i}}{\mu_{i}}}{2}}-1$, $i\in(b,e)$ and $z\in(\lambda,\hbar)$ where (b, $\lambda$) and (e, $\hbar$) corresponds to $\mathcal{N}-\mathcal{P}$ and $\mathcal{N}-\mathcal{Q}$ link, respectively. $\alpha_{i}$ and $\mu_{i}$ are two fading parameters that defines nonlinearity and clustering of the second hop \cite{juel2021secrecy}. Here, $\alpha_{i}$ is a positive integer \cite{lei2017secrecy} and $\varrho_{i}$ denotes the average SNR of this hop.


\subsection{PDFs for S-P and S-Q link}

Considering the instantaneous SNRs of $\mathcal{S}-\mathcal{P}$ and $\mathcal{S}-\mathcal{Q}$ link by $\xi_{sb}$ and $\xi_{se}$, respectively, their PDFs can be given as \cite[eq.~22,]{badrudduza2020enhancing},
\begin{align}
\label{9}
f_{\xi_{{s}{b}}}{(\xi)}=\frac{d{F_{\xi_{{sb}}}(\xi)}}{d{(\xi)}},
\\
\label{10}
f_{\xi_{{s}{e}}}{(\xi)}=\frac{d{F_{\xi_{{se}}}(\xi)}}{d{(\xi)}},
\end{align}
where $F_{\xi_{{sb}}}{(\xi)}$ and $F_{\xi_{{se}}}{(\xi)}$ are the Cumulative Distributive Functions (CDFs) of ${\xi_{{sb}}}$ and ${\xi_{{se}}}$, respectively. Here, $F_{\xi_{{sb}}}{(\xi)}$ can be defined as \cite[eq.~12,]{kumar2015performance}
\begin{align}
\label{11}
F_{\xi_{{sb}}}{(\xi)}=1-P{({\xi_{sr}}>{\xi_{sb}})}P{({\xi_{rb}}>{\xi_{sb}})},
\end{align}
where $P{({\xi_{sr}}>{\xi_{sb}})}$ and $P{({\xi_{rb}}>{\xi_{sb}})}$ represents the Complementary CDFs (CCDFs) of ${\xi_{sr}}$ and ${\xi_{rb}}$, respectively. The CCDF of ${\xi_{sr}}$ can be defined as \cite[eq.~23,]{badrudduza2020enhancing},
\begin{align}
\label{12}
P{({\xi_{sr}}>{\xi_{sb}})}=\int_{\xi_{sb}}^{\infty} f_{\xi_{{sr}}}{(\xi)}{d{\xi}}.
\end{align}
Substituting \eqref{7} into \eqref{12} and performing the integration using the identity \cite[eq.~3.351.2,]{GR:07:Book}, we get
\begin{align}
\label{13}
P{({\xi_{sr}}>{\xi_{sb}})}=\sum_{\theta_{1}=0}^{\infty}\psi_{s} {\chi_{s}}^{{-{\theta_{1}}-1}} \Gamma{({{\theta_{1}}+1},{\chi_{s}}{\xi})},
\end{align}
where $\Gamma(.,.)$ is a incomplete gamma function.

Similarly, the CCDF of ${\xi_{rb}}$ can be defined as
\begin{align}
\label{14}
P{({\xi_{rb}}>{\xi_{sb}})}=\int_{\xi_{sb}}^{\infty} f_{\xi_{{rb}}}{(\xi)}{d{\xi}}.
\end{align}
Now, substituting \eqref{8} into \eqref{14} and applying the identity of \cite[eq.~3.381.9,]{GR:07:Book},we get
\begin{align}
\label{15}
P{({\xi_{rb}}>{\xi_{sb}})}={\lambda_{3}}\Gamma{({\mu_{b}},{\psi_{1}}{{\xi}^{\frac{\alpha_{b}}{2}}})},
\end{align}
where $\lambda_{3}=\frac{2\lambda_{1}}{\alpha_{b}{\psi_1}^{\mu_{b}}}$. Using \eqref{13} and \eqref{15} into \eqref{11}, finally we will get $F_{\xi_{{sb}}}{(\xi)}$ as
\begin{align}
\label{16}
F_{\xi_{{sb}}}{(\xi)} =1-\sum_{\theta_{1}=0}^{\infty}\lambda_{{4},\theta_{1}} \Gamma{({{\theta_{1}}+1},{\chi_{s}}{\xi})}\Gamma{({\mu_{b}},{\psi_{1}}{{\xi}^{\frac{\alpha_{b}}{2}}})},
\end{align}
where $\lambda_{{4},\theta_{1}}=\psi_{s}{\chi_{s}}^{{-{\theta_{1}}-1}}\lambda_{3}$. Now, substituting \eqref{16} into \eqref{9} and differentiating this with respect to $\xi$, the PDF of $\xi_{{s}{b}}$ will be found as
\begin{align}
\label{17}
f_{\xi_{{s}{b}}}{(\xi)}&={\sum_{\theta_{1}=0}^{\infty}}{\lambda_{{4},{\theta_{1}}}}\biggl[\lambda_{5}{{\xi}^{\lambda_{2}}}{e^{-{\psi_{1}}{\xi^{\frac{\alpha_{b}}{2}}}}}{\Gamma{({{\theta_{1}}+1},{\chi_{s}}{\xi})}}+\lambda_{6}{{\xi}^{\theta_{1}}}{e^{-{\chi_{s}}{\xi}}}{\Gamma{({\mu_{b}},{\psi_{1}}{\xi^{\frac{\alpha_{b}}{2}}})}}\biggl],
\end{align}
where $\lambda_{5}=\frac{{{\psi_{1}}^{\mu_{b}}}\alpha_{b}}{2}$ and $\lambda_{6}={\chi_{s}}^{{\theta_{1}}+1}$.
Similarly the CDF of $\xi_{se}$ can be defined as \cite[eq.~12,]{kumar2015performance}
\begin{align}
\label{18}
F_{\xi_{{se}}}{(\xi)}=1-P{({\xi_{sr}}>{\xi_{se}})}P{({\xi_{re}}>{\xi_{se}})},
\end{align}
where $P{({\xi_{sr}}>{\xi_{se}})}$ and $P{({\xi_{re}}>{\xi_{se}})}$ represents the CCDFs ${\xi_{sr}}$ and ${\xi_{re}}$, respectively. Now by following the same procedure used in deriving $F_{\xi_{{sb}}}{(\xi)}$, the $F_{\xi_{{se}}}{(\xi)}$ can be given by
\begin{align}
\label{19}
F_{\xi_{{s}{e}}}{(\xi)}=1-\sum_{\theta_{2}=0}^{\infty}\hbar_{{4},\theta_{2}}{\Gamma{({{\theta_{2}}+1},{\chi_{s}}{\xi})}}\Gamma{({\mu_{e}},{\psi_{1}}{{\xi}^{\frac{\alpha_{e}}{2}}})},
\end{align}
where $\hbar_{{4},\theta_{2}}=\psi_{s} {\chi_{s}}^{-\theta_{2}-1}\hbar_{3}$ and $\hbar_{3}=\frac{2\hbar_{1}}{\alpha_{e}{\psi_1}^{\mu_{e}}}$. Substituting \eqref{19} into \eqref{10}, we get
\begin{align}
\label{20}
f_{\xi_{{s}{e}}}{(\xi)}&={\sum_{\theta_{2}=0}^{\infty}}{\hbar_{{4},{\theta_{2}}}}\biggl[\hbar_{5}{{\xi}^{\hbar_{2}}}{e^{-{\psi_{1}}{\xi^{\frac{\alpha_{e}}{2}}}}}{\Gamma{({{\theta_{2}}+1},{\chi_{s}}{\xi})}}+\hbar_{6}{{\xi}^{\theta_{2}}}{e^{-{\chi_{s}}{\xi}}}{\Gamma{({\mu_{e}},{\psi_{1}}{\xi^{\frac{\alpha_{e}}{2}}})}}\biggl],
\end{align}
where $\hbar_{5}=\frac{{{\psi_{1}}^{\mu_{e}}}\alpha_{e}}{2}$ and $\hbar_{6}={\chi_{s}}^{{\theta_{2}}+1}$.


\subsection{Best Relay Selection Procedure}


\subsubsection{Best Relay to $b$-th User}

Let, $\xi_{db}$ denotes the instantaneous SNR between the best relay and $b$th user. Now applying order statistics, the expression of $\xi_{db}$ can be shown as \cite{duong2009performance}
\begin{align}
\xi_{db}=arg ^{max}_{p\epsilon \mathcal{T}}min(\xi_{sr},\xi_{rb}).
\end{align}
The CDF of $\xi_{db}$ can be shown as
\begin{align}
\label{22}
F_{\xi_{{db}}}{(\xi)}={[F_{\xi_{{sb}}}{(\xi)}]}^{N}.
\end{align}
Substituting \eqref{16} into \eqref{22}, we get
\begin{align}
\label{23a}
F_{\xi_{{db}}}{(\xi)}={\biggl[1-\sum_{\theta_{1}=0}^{\infty}\lambda_{{4},\theta_{1}}{\Gamma{({{\theta_{1}}+1},{\chi_{s}}{\xi})}}\Gamma{({\mu_{b}},{\psi_{1}}{{\xi}^{\frac{\alpha_{b}}{2}}})}\biggl]}^{N}.
\end{align}
Differentiating \eqref{23a}, we get the PDF of ${\xi_{db}}$ as
\begin{align}
\nonumber
f_{\xi_{{db}}}{(\xi)}&={{\biggl[1-\sum_{\theta_{1}=0}^{\infty}\lambda_{{4},\theta_{1}}{\Gamma{({{\theta_{1}}+1},{\chi_{s}}{\xi})}}\Gamma{({\mu_{b}},{\psi_{1}}{{\xi}^{\frac{\alpha_{b}}{2}}})}\biggl]}^{N-1}} N\Bigg[{\sum_{\theta_{1}=0}^{\infty}}{\lambda_{{4},{\theta_{1}}}}\biggl(\lambda_{5}{{\xi}^{\lambda_{2}}}{e^{-{\psi_{1}}{\xi^{\frac{\alpha_{b}}{2}}}}}{\Gamma{({{\theta_{1}}+1},{\chi_{s}}{\xi})}}
\\
\label{24}
&
+\lambda_{6}{{\xi}^{\theta_{1}}}{e^{-{\chi_{s}}{\xi}}}{\Gamma{({\mu_{b}},{\psi_{1}}{\xi^{\frac{\alpha_{b}}{2}}})}}\biggl)\Bigg].
\end{align}


\subsubsection{Best Relay to $e$-th Eavesdropper}

Lets consider the instantaneous SNR between the best relay and $e$th eavesdropper as $\xi_{de}$. It can be similarly expressed as \cite{duong2009performance},
\begin{align}
\xi_{de}=arg ^{max}_{p\epsilon \mathcal{T}}min(\xi_{sr},\xi_{re}).
\end{align}
The CDF of $F_{\xi_{{de}}}{(\xi)}$ is given by,
\begin{align}
\label{26a}
F_{\xi_{{de}}}{(\xi)}={[F_{\xi_{{se}}}{(\xi)}]}^{N}.
\end{align}
Substituting \eqref{19} into \eqref{26a}, the CDF of ${\xi_{db}}$ can be written as
\begin{align}
\label{27a}
F_{\xi_{{de}}}{(\xi)}={\biggl[1-\sum_{\theta_{2}=0}^{\infty}\hbar_{{4},\theta_{2}}{\Gamma{({{\theta_{2}}+1},{\chi_{s}}{\xi})}}\Gamma{({\mu_{e}},{\psi_{1}}{{\xi}^{\frac{\alpha_{e}}{2}}})}\biggl]}^{N}.
\end{align}
Differentiating \eqref{27a}, we get the PDF of ${\xi_{de}}$ given by
\begin{align}
\nonumber
f_{\xi_{{de}}}{(\xi)}&={\biggl[1-\sum_{\theta_{2}=0}^{\infty}\hbar_{{4},\theta_{2}}{\Gamma{({{\theta_{2}}+1},{\chi_{s}}{\xi})}}\Gamma{({\mu_{e}},{\psi_{1}}{{\xi}^{\frac{\alpha_{e}}{2}}})}\biggl]^{N-1}} N\Bigg[{\sum_{\theta_{2}=0}^{\infty}}{\hbar_{{4},{\theta_{2}}}}\biggl(\hbar_{5}{{\xi}^{\hbar_{2}}}{e^{-{\psi_{1}}{\xi^{\frac{\alpha_{e}}{2}}}}}{\Gamma{({{\theta_{2}}+1},{\chi_{s}}{\xi})}}
\\
\label{28}
&+\hbar_{6}{{\xi}^{\theta_{2}}}{e^{-{\chi_{s}}{\xi}}}{\Gamma{({\mu_{e}},{\psi_{1}}{\xi^{\frac{\alpha_{e}}{2}}})}}\biggl)\Bigg].
\end{align}


\subsection{PDF of Multicast Channels}
We consider the worst case for the multicast channels (i.e. minimum SNR among $P$ users) because if the information at the weakest user's terminal is secure, then the total system can be considered secure. 
The minimum SNR of the multicast channels is denoted by $\zeta_{min}=min_{1<b<P}\xi_{db}$, and the corresponding PDF can be represented by 
\begin{align}
\label{29a}
f_{\zeta_{min}}(\xi)=P {f_{\xi_{db}}(\xi)}[1-{F_{\xi_{db}}(\xi)}]^{P-1}.
\end{align}
Putting values of \eqref{23a} and \eqref{24} into \eqref{29a}, we get
\begin{align}
\nonumber
f_{\zeta_{min}}(\xi)&={\sum_{\theta_{1}=0}^{\infty}}{\lambda_{{4},{\theta_{1}}}}{{\biggl[1-\sum_{\theta_{1}=0}^{\infty}\lambda_{{4},\theta_{1}}{\Gamma{({{\theta_{1}}+1},{\chi_{s}}{\xi})}}\Gamma{({\mu_{b}},{\psi_{1}}{{\xi}^{\frac{\alpha_{b}}{2}}})}\biggl]}^{N-1}}\Bigg[\biggl({\frac{\lambda_{5}{{\xi}^{\lambda_{2}}}}{e^{{\psi_{1}}{\xi^{\frac{\alpha_{b}}{2}}}}}}{\Gamma{({{\theta_{1}}+1},{\chi_{s}}{\xi})}}+{\frac{\lambda_{6}{{\xi}^{\theta_{1}}}}{e^{{\chi_{s}}{\xi}}}}{\Gamma{({\mu_{b}},{\psi_{1}}{\xi^{\frac{\alpha_{b}}{2}}})}}\biggl)\Bigg]
\\
\label{30}
&\times PN{\Biggl[{1-{\biggl[1-\sum_{\theta_{1}=0}^{\infty}\lambda_{{4},\theta_{1}}{\Gamma{({{\theta_{1}}+1},{\chi_{s}}{\xi})}}\Gamma{({\mu_{b}},{\psi_{1}}{{\xi}^{\frac{\alpha_{b}}{2}}})}\biggl]}^{N}\Biggl]^{P-1}}}.
\end{align}
\vspace{0mm}
Using the Identity of \cite[eq. 1.111]{GR:07:Book}, \eqref{30} can be written as 
\begin{align}
\nonumber
f_{\zeta_{min}}(\xi)&=PN\sum_{\theta_{1}=0}^{\infty}{\sum_{\theta_{3}=0}^{P-1}}{\lambda_{{4},{\theta_{1}}}}{\lambda_{{7},\theta_{3}}}{{\biggl[1-\sum_{\theta_{1}=0}^{\infty}\lambda_{{4},\theta_{1}}{\Gamma{({{\theta_{1}}+1},{\chi_{s}}{\xi})}}\Gamma{({\mu_{b}},{\psi_{1}}{{\xi}^{\frac{\alpha_{b}}{2}}})}\biggl]}^{N{(\theta_{3}+1)}-1}}
\\
\label{31}
&\times \biggl({\frac{\lambda_{5}{{\xi}^{\lambda_{2}}}}{e^{{\psi_{1}}{\xi^{\frac{\alpha_{b}}{2}}}}}}{\Gamma{({{\theta_{1}}+1},{\chi_{s}}{\xi})}}+{\frac{\lambda_{6}{{\xi}^{\theta_{1}}}}{e^{{\chi_{s}}{\xi}}}}{\Gamma{({\mu_{b}},{\psi_{1}}{\xi^{\frac{\alpha_{b}}{2}}})}}\biggl),
\end{align}
where ${\lambda_{{7},\theta_{3}}}={\binom{P-1}{{\theta_{3}}}}{(-1)^{\theta_{3}}}$. Equation (\ref{31}) can be further
simplified using \cite[eq.~8.352.7,]{GR:07:Book} and can be given by,
\begin{align}
\label{32}
f_{\zeta_{min}}(\xi)&=PN\sum_{\theta_{1}=0}^{\infty}{\sum_{\theta_{3}=0}^{P-1}}{\sum_{{\theta_{4}}=0}^{N{(\theta_{3}+1)}-1}}{\lambda_{{4},{\theta_{1}}}}{\lambda_{{7},\theta_{3}}}{\lambda_{{8},\theta_{4}}}\Biggl[{{\sum_{\theta_{1}=0}^{\infty}}{\sum_{\theta_{5}=0}^{\theta_{1}}}{\sum_{\theta_{6}=0}^{\mu_{b}-1}}}{\delta_{1}}{(\xi)}\Biggl]^{\theta_{4}}\biggl({\frac{\lambda_{5}{{\xi}^{\lambda_{2}}}}{e^{{\psi_{1}}{\xi^{\frac{\alpha_{b}}{2}}}}}}{\Gamma{({{\theta_{1}}+1},{\chi_{s}}{\xi})}}+{\frac{\lambda_{6}{{\xi}^{\theta_{1}}}}{e^{{\chi_{s}}{\xi}}}}{\Gamma{({\mu_{b}},{\psi_{1}}{\xi^{\frac{\alpha_{b}}{2}}})}}\biggl),
\end{align}
where 
$\delta_{1}(\xi)=\lambda_{{9},\theta_{{1},{5},{6}}}\xi^{\lambda_{{10},{\theta_{{5},{6}}}}}e^{-{{\chi_{s}}\xi}-{{\psi_{1}}\xi^{\frac{\alpha_{b}}{2}}}}$, ${\lambda_{{10},\theta_{{5},{6}}}}=\theta_{5}+{\frac{\alpha_{b}}{2}}{\theta_{6}}$, ${\lambda_{{9},\theta_{{1},{5},{6}}}}=\frac{{\lambda_{{4},\theta_{1}}{(\theta_{1})!}}{{\chi_{s}}^{\theta_{5}}}{(\mu_{b}-1)!}{\psi_{1}}^{\theta_{6}}}{{\theta_{5}!}{\theta_{6}!}}$, 
and ${\lambda_{{8},\theta_{4}}}={(-1)^{\theta_{4}}}$
\\
${\binom{{N{(\theta_{3}+1)}-1}}{{\theta_{4}}}}$.
The solution of the term $\big[{\sum_{\theta_{1}=0}^{\infty}}$ ${\sum_{\theta_{5}=0}^{\theta_{1}}} $ ${\sum_{\theta_{6}=0}^{{\mu_{b}}-1}} $ ${\delta_{1}}{(\xi)}\big]^{\theta_{4}}$ given in \eqref{32} requires a special mathematical tool named multinomial theorem. According to this theorem, $\big[{\sum_{\theta_{1}=0}^{\infty}}{\sum_{\theta_{5}=0}^{\theta_{1}}}{\sum_{\theta_{6}=0}^{{\mu_{b}}-1}}{\delta_{1}}{(\xi)}\big]^{\theta_{4}}$ can be represented by \cite[eq.~7,]{pena2014performance},
\begin{align}
\label{33}
\biggl[{\sum_{\theta_{1}=0}^{\infty}}{\sum_{\theta_{5}=0}^{\theta_{1}}}{\sum_{\theta_{6}=0}^{{\mu_{b}}-1}}{\delta_{1}}{(\xi)}\biggl]^{\theta_{4}}&={\sum_{\tau_{\theta_{4}}}}{{\Lambda_{1,{\tau_{\theta_{4}}}}}}{\xi^{{\Lambda_{4,{\tau_{\theta_{4}}}}}}}{e^{{-{\Lambda_{2,{\tau_{\theta_{4}}}}{\xi}}}{-{\Lambda_{3,{\tau_{\theta_{4}}}},{\xi^{\frac{\alpha_{b}}{2}}}}}}},
\end{align}
where 
${{\Lambda_{1,{\tau_{\theta_{4}}}}}}={{\Lambda_{5,{\tau_{\theta_{4}}}}}}{{{\binom{\theta_{4}}{{t_{{0},{0},{0}}},...,{t_{\theta_{1},\theta_{5},\theta_{6}}},...,{t_{{{m_{s}}-1},{\theta_{1}},{{\mu_{b}}-1}}}}}}}$,
${{\Lambda_{5,{\tau_{\theta_{4}}}}}}$ $={\Pi_{{\theta_{1}},{\theta_{5}},{\theta_{6}}}}{(\lambda_{{9},\theta_{{1},{5},{6}}})^{t_{{\theta_{1}},{\theta_{5}},{\theta_{6}}}}}$, $\binom{u}{u_{1}, u_{2},...,u_{\theta}}=\frac{u!}{u_{1}!, u_{2}!,...,u_{\theta}!}$, 
\\
${{\Lambda_{3,{\tau_{\theta_{4}}}}}}= {\sum_{\theta_{1}}}{\sum_{\theta_{5}}}{\sum_{\theta_{6}}
{\psi_{1}}{t_{{\theta_{1}},{\theta_{5}},{\theta_{6}}}}}$,
$\Lambda_{4,{\tau_{\theta_{4}}}}=\sum_{\theta_{1}}$ $\sum_{\theta_{5}}$  $\sum_{\theta_{6}}$ $\lambda_{10,\theta_{5,6}}t_{\theta_{1},{\theta_{5}},{\theta_{6}}}$,
and ${{\Lambda_{2,{\tau_{\theta_{4}}}}}}= {\sum_{\theta_{1}}}{\sum_{\theta_{5}}}{\sum_{\theta_{6}}}$,
${{\chi_{s}}{t_{{\theta_{1}},{\theta_{5}},{\theta_{6}}}}}$ describes the multinomial coefficients. For every element of $\tau_{\theta_{4}}$, the sum in \eqref{33} can be given by
\begin{align}
\label{34}
\tau_{\theta_{4}}=&\Big[{t_{{0},{0},{0}}},...,{t_{\theta_{1},\theta_{5},\theta_{6}}},...,{t_{{{m_{s}}-1},{\theta_{1}},{{\mu_{b}}-1}}}:{t_{\theta_{1},\theta_{5},\theta_{6}}}\in \mathbb{N}, 0\leq{\theta_{1}}\leq{\infty}, 0\leq{\theta_{5}}\leq{\theta_{1}}, {0\leq{\theta_{6}}\leq{\mu_{b}-1}};\sum_{\theta_{1}, \theta_{5}, \theta_{6}} {t_{\theta_{1}, \theta_{5}, \theta_{6}}}\Big].
\end{align}
Now, substituting \eqref{33} into \eqref{32}, we get
\begin{align}
\label{35}
f_{\zeta_{min}}(\xi)&= \sum_{\theta_{1}=0}^{\infty}{\sum_{\theta_{3}=0}^{P-1}}{\sum_{{\theta_{4}}=0}^{N{(\theta_{3}+1)}-1}}{\sum_{\tau_{\theta_{4}}}}\frac{\lambda_{{11},\theta_{{1},{3},{4}}}}{e^{{{\lambda_{14}}\xi}+{\lambda_{15}}\xi^{\frac{\alpha_{b}}{2}}}}{\Biggl[{\sum_{\theta_{7}=0}^{\theta_{1}}{\lambda_{{12},\theta_{7}}}\xi^{r_1}+{\sum_{\theta_{8}=0}^{\mu_{b}-1}{\lambda_{{13},\theta_{8}}}\xi^{r_2}}\Biggl]}},
\end{align}
where $\lambda_{{11},\theta_{{1},{3},{4}}}=PN{\lambda_{{4},{\theta_{1}}}}{\lambda_{{7},\theta_{3}}}{\lambda_{{8},\theta_{4}}}{{\Lambda_{1,{\tau_{\theta_{4}}}}}}$, $r_{1}={\Lambda_{4}},{\tau_{\theta_{4}}}+{{\lambda_{2}}+{\theta_{7}}}$, $\lambda_{{13},\theta_{8}}=\frac{{\lambda_{6}}{({\mu_{b}}-1)!{{\psi_{1}}^{\theta_{8}}}}}{\theta_{8}}$,
$\lambda_{14}=\chi_{s}+\Lambda_{2,{\tau_{\theta_{4}}}}$,  $\lambda_{15}=\psi_{1}+\Lambda_{3,{\tau_{\theta_{4}}}}$, $\lambda_{{12},\theta_{7}}=\frac{{\lambda_{5}}{(\theta_{1})!{{\chi_{s}}^{\theta_{7}}}}}{\theta_{7}}$ and $r_{2}={\Lambda_{4}},{\tau_{\theta_{4}}}+{{\theta_{1}}+\frac{{\alpha_{b}}{\theta_{8}}}{2}}$.


\subsection{PDF of Eavesdropper Channel}
To ensure a perfect secrecy, we therein consider the strongest case of the eavesdropper channels (i.e. maximum SNR among $Q$ eavesdroppers). The reason is that if the transmitted information can be protected from the strongest eavesdropper, then the total system can be considered secure from the all remaining eavesdroppers. Hence, the maximum SNR of eavesdropper channels is denoted by $\zeta_{max}=max_{1<e<Q}\xi_{de}$ and its PDF can be represented by
\begin{align}
\label{36}
f_{\zeta_{max}}(\xi) =Q f_{\xi_{de}}(\xi)[F_{\xi_{de}}(\xi)]^{Q-1}.
\end{align}
Putting values of \eqref{27a} and \eqref{28} into \eqref{36}, we get,
\begin{align}
\label{37}
f_{\zeta_{max}}(\xi)&=QN{\Bigg[{\sum_{\theta_{2}=0}^{\infty}}{\hbar_{{4},{\theta_{2}}}}\biggl([\hbar_{5}{{\xi}^{\hbar_{2}}}{e^{-{\psi_{1}}{\xi^{\frac{\alpha_{e}}{2}}}}}{\Gamma{({{\theta_{2}}+1},{\chi_{s}}{\xi})}}\Bigg]}\biggl[1-\sum_{\theta_{2}=0}^{\infty}\hbar_{{4},\theta_{2}}{\Gamma{({{\theta_{2}}+1},{\chi_{s}}{\xi})}}\Gamma{({\mu_{e}},{\psi_{1}}{{\xi}^{\frac{\alpha_{e}}{2}}})}\biggl]^{QN-1}.
\end{align}
Now applying the Identity of \cite[eq. 1.111]{GR:07:Book} and \cite[ eq.8.352.7]{GR:07:Book}, \eqref{37} can be written as
\begin{align}
\label{38}
f_{\zeta_{max}}(\xi)&=\sum_{\theta_{2}=0}^{\infty}{\sum_{\theta_{11}=0}^{QN-1}}{\hbar_{{4},{\theta_{2}}}}{\hbar_{{7},\theta_{11}}}\Biggl[{{\sum_{\theta_{2}=0}^{\infty}}{\sum_{\theta_{9}=0}^{\theta_{2}}}{\sum_{\theta_{10}=0}^{\mu_{e}-1}}}{\delta_{2}}{(\xi)}\Biggl]^{\theta_{11}}e^{-{{\chi_{s}}\xi}-{\psi_{1}}\xi^{\frac{\alpha_{e}}{2}}}{\Biggl[{\sum_{\theta_{12}=0}^{\theta_{2}}{\hbar_{{9},\theta_{12}}}\xi^{{\hbar_{2}}+{\theta_{12}}}}+{\sum_{\theta_{13}=0}^{\mu_{e}-1}{\hbar_{{10},\theta_{13}}}\xi^{({\theta_{2}}+\frac{{\alpha_{e}}{\theta_{13}}}{2})}}\Biggl]},
\end{align}
where 
${\delta_{2}}{(\xi)}=\hbar_{{8},\theta_{{2},{9},{10}}}{\xi^{\theta_{9}+\frac{{\alpha_{e}}{\theta_{10}}}{2}}}e^{-{{\chi_{s}}\xi}-{{\psi_{1}}\xi^{\frac{\alpha_{e}}{2}}}}$, ${\hbar_{{7},\theta_{11}}}=$ ${\binom{QN-1}{{\theta_{11}}}}$ $QN$, ${\hbar_{{8},\theta_{{2},{9},{10}}}}= \frac{{(-1)}{\hbar_{{4},\theta_{2}}{(\theta_{2})!}}{{\chi_{s}}^{\theta_{9}}}{(\mu_{e}-1)!}{\psi_{1}}^{\theta_{10}}}{{\theta_{9}!}{\theta_{10}!}}$, 
$\hbar_{{9},\theta_{12}}= \frac{{\hbar_{5}}{(\theta_{2})!}}{{(\theta_{12})}!}$ ${{\chi_{s}}^{\theta_{12}}}$, 
$\hbar_{{10},\theta_{13}}= \frac{{\hbar_{6}}{({\mu_{e}}-1)!{{\psi_{1}}^{\theta_{13}}}}}{{(\theta_{13})}!}$ 
and 
${\hbar_{{11},\theta_{{9},{10}}}}={\theta_{9}+\frac{{\alpha_{e}}{\theta_{10}}}{2}}$. Again,
$\Biggl[{{\sum_{\theta_{2}=0}^{\infty}}{\sum_{\theta_{9}=0}^{\theta_{2}}}{\sum_{\theta_{10}=0}^{\mu_{e}-1}}}{\delta_{2}}{(\xi)}\Biggl]^{\theta_{11}}$ can be represented using the Multinomial theorem as given by
\begin{align}
\label{39}
\Biggl[{{\sum_{\theta_{2}=0}^{\infty}}{\sum_{\theta_{9}=0}^{\theta_{2}}}{\sum_{\theta_{10}=0}^{\mu_{e}-1}}}{\delta_{2}}{(\xi)}\Biggl]^{\theta_{11}}={\sum_{\tau_{\theta_{11}}}}{\Lambda_{6},{\tau_{\theta_{11}}}}{e^{{-{\Lambda_{7},{{\tau_{\theta_{11}}}\xi}}}{-{\Lambda_{8},{{\tau_{\theta_{11}}}\xi^{\frac{\alpha_{e}}{2}}}}}}}{\xi^{{\Lambda_{9}},{\tau_{\theta_{11}}}}},
\end{align}
where 
${\Lambda_{6},{\tau_{\theta_{11}}}}={\Lambda_{10},{\tau_{\theta_{11}}}}{{{\binom{\theta_{11}}{{t_{{0},{0},{0}}},...,{t_{\theta_{2},\theta_{9},\theta_{10}}},...,{t_{{{m_{s}}-1},{\theta_{8}},{{\mu_{e}}-1}}}}}}}$, ${\Lambda_{10},{\tau_{\theta_{11}}}}={\Pi_{{\theta_{2}},{\theta_{9}},{\theta_{10}}}}{(\hbar_{{8},\theta_{{2},{8},{9}}})^{t_{{\theta_{2}},{\theta_{9}},{\theta_{10}}}}}$, 
${\Lambda_{7},{\tau_{\theta_{11}}}}= $ ${\sum_{\theta_{2}}}$ ${\sum_{\theta_{9}}}$ ${\sum_{\theta_{10}}{\chi_{s}}{t_{{\theta_{2}},{\theta_{9}},{\theta_{10}}}}}$,
${\Lambda_{8},{\tau_{\theta_{11}}}}= {\sum_{\theta_{2}}}{\sum_{\theta_{9}}}{\sum_{\theta_{10}}{\psi_{1}}{t_{{\theta_{2}},{\theta_{9}},{\theta_{10}}}}}$ 
and
${\Lambda_{9},{\tau_{\theta_{11}}}}= {\sum_{\theta_{2}}}{\sum_{\theta_{9}}}{\sum_{\theta_{10}}{\lambda_{{11},{\theta_{{9},{10}}}}}{t_{{\theta_{2}},{\theta_{9}},{\theta_{10}}}}}$, where
\begin{align}
\label{40}
\tau_{\theta_{11}}=&\Big[{t_{{0},{0},{0}}},...,{t_{\theta_{2},\theta_{9},\theta_{10}}},...,{t_{{{m_{s}}-1},{\theta_{2}},{{\mu_{e}}-1}}}:{t_{\theta_{2},\theta_{9},\theta_{10}}}\in \mathbb{N}, 0\leq{\theta_{2}}\leq{{\infty}, 0\leq{\theta_{9}}\leq{\theta_{2}}, {0\leq{\theta_{10}}\leq{\mu_{e}-1}}};\sum_{\theta_{2}, \theta_{9}, \theta_{10}} {t_{\theta_{2}, \theta_{9}, \theta_{10}}}\Big].
\end{align}
Substituting \eqref{39} into \eqref{38}, we get
\begin{align}
\label{41}
f_{\zeta_{max}}(\xi)&=\sum_{\theta_{2}=0}^{\infty}{\sum_{\theta_{11}=0}^{QN-1}}{\sum_{\tau_{\theta_{11}}}}{\hbar_{12,\theta_{2,11}}}e^{-{{\hbar_{13}}\xi}-{\hbar_{14}}\xi^{\frac{\alpha_{e}}{2}}} \Biggl[{\sum_{\theta_{12}=0}^{\theta_{2}}{\hbar_{{9},\theta_{12}}}\xi^{r_3}}+{\sum_{\theta_{13}=0}^{\mu_{e}-1}{\hbar_{{10},\theta_{13}}}\xi^{r_4}\Biggl]},
\end{align}
where
$\hbar_{{12},\theta_{{2},{11}}}={\hbar_{{4},{\theta_{2}}}}{\hbar_{{7},\theta_{11}}}{\Lambda_{6},{\tau_{\theta_{11}}}}$, $\hbar_{13}=\chi_{s}+\Lambda_{7},{\tau_{\theta_{11}}}$, $\hbar_{14}=\psi_{1}+\Lambda_{8},{\tau_{\theta_{11}}}$, $r_{3}={\Lambda_{9}},{\tau_{\theta_{11}}}+{{\hbar_{2}}+{\theta_{12}}}$ and 
$r_{4}={\Lambda_{9}},{\tau_{\theta_{11}}}+{{\theta_{2}}+\frac{{\alpha_{e}}{\theta_{13}}}{2}}$.


\section{SOPM Analysis}
\label{sec3}

We denote secrecy multicast capacity as $C_{s}$ \cite{wyner1975wire} and the target secrecy rate as $\varphi_{c}$. The SOPM indicates the transmitted information will be secured only if $C _{s}>\varphi_{c}$. Mathematically, the SOPM is defined as \cite[eq.~20,]{9139494}
\begin{align}
\label{42}
P_{out}(\varphi_{c})&=1-\int_{0}^{\infty}\left(\int_{w}^{\infty}f_{\zeta_{min}}(\xi_{db})d{\xi_{db}}\right)f_{\zeta_{max}}(\xi_{de})d{\xi_{de}}=1-\int_{0}^{\infty}\mathcal{I}_{1}f_{\zeta_{max}}(\xi_{de})d{\xi_{de}},
\end{align}
where $w=2^{\varphi_{c}}{(1+{\xi_{de}})-1}$, $\varphi_c>0$ and
$\mathcal{I}_{1}=$ $\int_{w}^{\infty}$ $f_{\zeta_{min}}$ $(\xi_{db})$ $d{\xi_{db}}$.

\subsection{Derivation of ${I}_{1}$}

$\mathcal{I}_{1}$ can be written with the help of \eqref{35} as
\begin{align}
\label{44}
\mathcal{I}_{1}&=\sum_{\theta_{1}=0}^{\infty}{\sum_{\theta_{3}=0}^{P-1}}{\sum_{{\theta_{4}}=0}^{N{(\theta_{3}+1)}-1}}{\sum_{\tau_{\theta_{4}}}}\lambda_{{11},\theta_{{1},{3},{4}}}\int_{w}^{\infty}{\Biggl[{\sum_{\theta_{7}=0}^{\theta_{1}}{\lambda_{{12},\theta_{7}}}{\xi_{db}}^{r_{1}}}+{\sum_{\theta_{8}=0}^{\mu_{b}-1}{\lambda_{{13},\theta_{8}}}{\xi_{db}}^{r_{2}}}\Biggl]} e^{-{{\lambda_{14}}{\xi_{db}}}}e^{-{\lambda_{15}}{\xi_{db}}^{\frac{\alpha_{b}}{2}}}d{\xi_{db}}.
\end{align}
Now, using the series representation of exponential function i.e. $e^{-\lambda_{15}\xi_{ab}^{\frac{\alpha_{b}}{2}}}=\sum_{\theta_{14}}^{\infty}\frac{(-\lambda_{15})^{\theta_{14}}\xi_{ab}^{\frac{\alpha_{b}\theta_{14}}{2}}}{\theta_{14}!}$ using the identity of \cite[eq.~1.211.1,] {GR:07:Book}, \eqref{44} can be simplified as
\begin{align}
\label{45}
\mathcal{I}_{1}&=\sum_{\theta_{1}=0}^{\infty}{\sum_{\theta_{3}=0}^{P-1}}\sum_{{\theta_{4}}=0}^{N{(\theta_{3}+1)}-1}\sum_{\tau_{\theta_{4}}}\sum_{\theta_{14}}^{\infty}\lambda_{{11},\theta_{{1},{3},{4},14}}e^{-{{\lambda_{14}}{\xi_{db}}}} \int_{w}^{\infty}{\Biggl[{\sum_{\theta_{7}=0}^{\theta_{1}}{\lambda_{{12},\theta_{7}}}{\xi_{db}}^{r_{5}}}+{\sum_{\theta_{8}=0}^{\mu_{b}-1}{\lambda_{{13},\theta_{8}}}{\xi_{db}}^{r_{6}}}\Biggl]}d{\xi_{db}},
\end{align}
where $\lambda_{{11},\theta_{{1},{3},{4},14}}=\lambda_{{11},\theta_{{1},{3},{4}}}\frac{(-\lambda_{15})^{\theta_{14}}}{\theta_{14}!}$, $r_{5}=r_{1}+\frac{\alpha_{b}\theta_{14}}{2}$, and $r_{6}=r_{2}+\frac{\alpha_{b}\theta_{14}}{2}$.
At first, integrating with the help of \cite[eq.~3.351.2,]{GR:07:Book}, and then using binomial series expansion of \cite[eq.~1.111,]{GR:07:Book}, $\mathcal{I}_{1}$ is derived as
\begin{align}
\label{46}
\mathcal{I}_{1}=\sum_{\theta_{1}=0}^{\infty}{\sum_{\theta_{3}=0}^{P-1}}\sum_{{\theta_{4}}=0}^{N{(\theta_{3}+1)}-1}\sum_{\tau_{\theta_{4}}}\sum_{\theta_{14}}^{\infty}\lambda_{{11},\theta_{{1},{3},{4},14}}e^{-\lambda_{14}\varphi_{2}\xi_{de}}\Biggl[
\sum_{\theta_{7}=0}^{\theta_{1}}\sum_{\theta_{15}}^{r_{5}}\sum_{\theta_{17}}^{\theta_{15}}\lambda_{{12},\theta_{7,15,17}}
\xi_{de}^{\theta_{17}}
+\sum_{\theta_{8}=0}^{\mu_{b}-1}\sum_{\theta_{16}}^{r_{6}}\sum_{\theta_{18}}^{\theta_{16}}\lambda_{{13},\theta_{8,16,18}} \xi_{de}^{\theta_{18}}
\Biggl],
\end{align}
where $\lambda_{{12},\theta_{7,15,17}}=\lambda_{{12},\theta_{7}}\binom{\theta_{15}}{\theta_{17}}\frac{r_{5}!e^{-\lambda_{14}\varphi_{1}}\varphi_{1}^{\theta_{15}-\theta_{17}}\varphi_{2}^{\theta_{17}}}{\theta_{15}!\lambda_{14}^{r_{5}-\theta_{15}+1}}$, $\varphi_{1}=2^{\varphi_{c}}-1$, $\varphi_{2}=2^{\varphi_{c}}$, and
$\lambda_{{13},\theta_{8,16,18}}=\binom{\theta_{16}}{\theta_{18}}\frac{r_{6}!e^{-\lambda_{14}\varphi_{1}}\varphi_{1}^{\theta_{16}-\theta_{18}}\varphi_{2}^{\theta_{18}}}{\theta_{16}!\lambda_{14}^{r_{6}-\theta_{16}+1}}$. Now, substituting the value of $\mathcal{I}_{1}$ into \eqref{42}, we get the final expression of SOPM as given in \eqref{sop}. 
\begin{align}
\nonumber
P_{out}(\varphi_{c})&=1- \sum_{\theta_{1}=0}^{\infty} {\sum_{\theta_{3}=0}^{P-1}} \sum_{{\theta_{4}}=0}^{N{(\theta_{3}+1)}-1} \sum_{\tau_{\theta_{4}}} \sum_{\theta_{14}}^{\infty} \sum_{\theta_{2}=0}^{\infty} {\sum_{\theta_{11}=0}^{QN-1}} \sum_{\tau_{\theta_{11}}} \lambda_{{11},\theta_{{1},{3},{4},14}}\hbar_{12,\theta_{2,11}} \Biggl[ \sum_{\theta_{7}=0}^{\theta_{1}}\sum_{\theta_{15}}^{r_{5}}\sum_{\theta_{17}}^{\theta_{15}} \Biggl(\sum_{\theta_{12}=0}^{\theta_{2}}{\hbar_{{9},\theta_{12}}}\lambda_{{12},\theta_{7,15,17}} \mathcal{I}_{2}
\\
\label{sop}
&+\sum_{\theta_{13}=0}^{\mu_{e}-1}{\hbar_{{10},\theta_{13}}}\lambda_{{12},\theta_{7,15,17}} \mathcal{I}_{3}\Biggl)+\sum_{\theta_{8}=0}^{\mu_{b}-1}\sum_{\theta_{16}}^{r_{6}}\sum_{\theta_{18}}^{\theta_{16}} \Biggl( \sum_{\theta_{12}=0}^{\theta_{2}}{\hbar_{{9},\theta_{12}}}\lambda_{{13},\theta_{8,16,18}} \mathcal{I}_{4}
+ \sum_{\theta_{13}=0}^{\mu_{e}-1}{\hbar_{{10},\theta_{13}}}\lambda_{{13},\theta_{8,16,18}} \mathcal{I}_{5}\Biggl)\Biggl],
\end{align}
where $r_5=r_3+\theta_{17}$, $r_6=r_4+\theta_{17}$, $r_7=r_3+\theta_{18}$, $r_8=r_4+\theta_{18}$ and $r_9=\lambda_{14}\varphi_{2}+\hbar_{13}$. Here the four integral terms (i.e. $\mathcal{I}_{2}, \mathcal{I}_{3}, \mathcal{I}_{4}$ and $\mathcal{I}_{5}$) can be derived as the following:


\subsection{Derivation of ${I}_{2}, {I}_{3}, {I}_{4}$, and ${I}_{5}$}
The expression of $\mathcal{I}_{2}$ is assumed as
\setcounter{equation}{46}
\begin{align}
\label{101}
\mathcal{I}_{2}&=\int_{0}^{\infty} \xi_{de}^{r_{5}} e^{-{ r_9 \xi_{de}}}e^{-{\hbar_{14}}\xi_{de}^{\frac{\alpha_{e}}{2}}} d\xi_{de}.
\end{align}
For mathematical simplicity, we can convert the exponential functions to the well known Meijer-$G$ functions which is an exceptional mathematical tool accepted and widely used by the researchers by applying the following identity \cite[eq.~8.4.3.1,]{prudnikov1988integrals}. 
\begin{align}
\nonumber
    e^{-x}=G_{0, 1}^{1, 0} \left[x\Biggl|\begin{array}{c}
-  \\
0
\end{array}
\right].
\end{align}
Applying this identity, \eqref{101} can be given as
\begin{align}
\label{102}
    \mathcal{I}_{2}=\int_{0}^{\infty} \xi_{de}^{r_{5}} G_{0, 1}^{1, 0} \left[r_{9}\xi_{de}\Biggl|\begin{array}{c}
-  \\
0
\end{array}
\right]
G_{0, 1}^{1, 0} \left[\hbar_{14}\xi_{de}^{\frac{\alpha_{e}}{2}}\Biggl|\begin{array}{c}
-  \\
0
\end{array}
\right]
d\xi_{de}.
\end{align}
Now, by performing simple integration of \eqref{102} using \cite[eq.~2.24.1.1,]{prudnikov1988integrals}, we get
\begin{align}
\label{103}
\mathcal{I}_{2}=\frac{\sqrt{2}\alpha_{e}^{\frac{1}{2}+r_{5}}}{r_{9}^{r_{5}+1}(2\pi)^{\frac{\alpha_{e}}{2}}}G_{\alpha_{e}, 2}^{2, \alpha_{e} }\left[\frac{\hbar_{14}^{2}2^{-2}}{r_{9}^{\alpha_{e}}\alpha_{e}^{-\alpha_{e}}}\Biggl|\begin{array}{c}
\Delta(\alpha_{e},-r_{5})  \\
\Delta(2,0)
\end{array}
\right],
\end{align}
where $\Delta(a,b)=\frac{b}{a}, \frac{b+1}{a}, \frac{b+2}{a},...,\frac{b+a-1}{a}$. Similarly, the expressions of  $\mathcal{I}_{3}, \mathcal{I}_{4}$ and $\mathcal{I}_{5}$ can be derived as follows.
\begin{align}
\nonumber
\mathcal{I}_{3}&=\int_{0}^{\infty} \xi_{de}^{r_6}e^{-{ r_9 \xi_{de}}}e^{-{\hbar_{14}}\xi_{de}^{\frac{\alpha_{e}}{2}}}d\xi_{de}
=\int_{0}^{\infty} \xi_{de}^{r_{6}} G_{0, 1}^{1, 0} \left[r_{9}\xi_{de}\Biggl|\begin{array}{c}
-  \\
0
\end{array}
\right]
G_{0, 1}^{1, 0} \left[\hbar_{14}\xi_{de}^{\frac{\alpha_{e}}{2}}\Biggl|\begin{array}{c}
-  \\
0
\end{array}
\right]
d\xi_{de}
\\
\label{104}
&
=\frac{\sqrt{2}\alpha_{e}^{\frac{1}{2}+r_{6}}}{r_{9}^{r_{6}+1}(2\pi)^{\frac{\alpha_{e}}{2}}}G_{\alpha_{e}, 2}^{2, \alpha_{e} }\left[\frac{\hbar_{14}^{2}2^{-2}}{r_{9}^{\alpha_{e}}\alpha_{e}^{-\alpha_{e}}}\Biggl|\begin{array}{c}
\Delta(\alpha_{e},-r_{6})  \\
\Delta(2,0)
\end{array}
\right].
\end{align}
\begin{align}
\nonumber
\mathcal{I}_{4}&=\int_{0}^{\infty} \xi_{de}^{r_7} e^{-{ r_9 \xi_{de}}}e^{-{\hbar_{14}}\xi_{de}^{\frac{\alpha_{e}}{2}}}d\xi_{de}
=\int_{0}^{\infty} \xi_{de}^{r_{7}} G_{0, 1}^{1, 0} \left[r_{9}\xi_{de}\Biggl|\begin{array}{c}
-  \\
0
\end{array}
\right]
G_{0, 1}^{1, 0} \left[\hbar_{14}\xi_{de}^{\frac{\alpha_{e}}{2}}\Biggl|\begin{array}{c}
-  \\
0
\end{array}
\right]
d\xi_{de}
\\
\label{105}
&=\frac{\sqrt{2}\alpha_{e}^{\frac{1}{2}+r_{7}}}{r_{9}^{r_{7}+1}(2\pi)^{\frac{\alpha_{e}}{2}}}G_{\alpha_{e}, 2}^{2, \alpha_{e} }\left[\frac{\hbar_{14}^{2}2^{-2}}{r_{9}^{\alpha_{e}}\alpha_{e}^{-\alpha_{e}}}\Biggl|\begin{array}{c}
\Delta(\alpha_{e},-r_{7})  \\
\Delta(2,0)
\end{array}
\right].
\end{align}
\begin{align}
\nonumber
\mathcal{I}_{5}&=\int_{0}^{\infty} \xi_{de}^{r_8} e^{-{ r_9 \xi_{de}}}e^{-{\hbar_{14}}\xi_{de}^{\frac{\alpha_{e}}{2}}}d\xi_{de}
=\int_{0}^{\infty} \xi_{de}^{r_{8}} G_{0, 1}^{1, 0} \left[r_{9}\xi_{de}\Biggl|\begin{array}{c}
-  \\
0
\end{array}
\right]
G_{0, 1}^{1, 0} \left[\hbar_{14}\xi_{de}^{\frac{\alpha_{e}}{2}}\Biggl|\begin{array}{c}
-  \\
0
\end{array}
\right]
d\xi_{de}
\\
\label{105}
&=\frac{\sqrt{2}\alpha_{e}^{\frac{1}{2}+r_{8}}}{r_{9}^{r_{8}+1}(2\pi)^{\frac{\alpha_{e}}{2}}}G_{\alpha_{e}, 2}^{2, \alpha_{e} }\left[\frac{\hbar_{14}^{2}2^{-2}}{r_{9}^{\alpha_{e}}\alpha_{e}^{-\alpha_{e}}}\Biggl|\begin{array}{c}
\Delta(\alpha_{e},-r_{8})  \\
\Delta(2,0)
\end{array}
\right].
\end{align}

\subsection{Generalization offered by the SOPM expression}
Since our proposed second hop is a generalized model, the final generalized expression of SOPM given in \eqref{sop} can be used to perfectly measure the probability of information outage of various classical multipath fading channels as our special cases. For example, this expression can directly replicate the results of \cite[eq.~25,]{bankey2017secrecy} for $P=1$, $Q=1$, $\alpha=2$, and $\mu=m$, and \cite[eq.~27,]{bankey2019physical} for $N=1$, $P=1$, $\alpha=2$, and $\mu=m$, where $m$ is the Nakagami-$m$ fading parameter.


\section{ESMC Analysis}
\label{sec4}

The ESMC signifies the average values of instantaneous secrecy capacity which can be presented as \cite[eq.~4,]{liu2015ergodic}
\begin{align}
\label{50}
\langle D_{o}^{mc}\rangle&=\int_{0}^{\infty}\log_{2}(1+\xi_{db})f_{\zeta_{min}}(\xi_{db})d{\xi_{db}}-\int_{0}^{\infty}\log_{2}(1+\xi_{de})f_{\zeta_{max}}(\xi_{de})d{\xi_{de}}=\Upsilon_{1}-\Upsilon_{2}.
\end{align}
\subsection{Derivation of $\Upsilon_{1}$}
The expression of $\Upsilon_{1}$ is presumed as
\begin{align}
\label{201}
    \Upsilon_{1}&=\int_{0}^{\infty}\log_{2}(1+\xi_{db})f_{\zeta_{min}}(\xi_{db})d{\xi_{db}}=\frac{1}{\ln(2)}\int_{0}^{\infty}\ln(1+\xi_{db})f_{\zeta_{min}}(\xi_{db})d{\xi_{db}}.
\end{align}
Again, for simplicity of mathematical formulation, $ln(1+\xi_{db})$ can be transformed into Meijer-$G$ function by employing the identity \cite[eq.~8.4.6.5,]{prudnikov1988integrals} as given by
\begin{align}
\label{202}
    \ln(1+\xi_{db})=G_{2, 2}^{1, 2} \left[1+\xi_{db}\Biggl|\begin{array}{c}
1, 1  \\
1, 0
\end{array}
\right].
\end{align}
Now, using \eqref{35}, and \eqref{202} in \eqref{201}, and integrating it applying the formulas of \cite[eq.~1.211.1,]{GR:07:Book}, and \cite[eq.~8.4.3.1,]{prudnikov1988integrals}, $\Upsilon_{1}$ can be written as
\begin{align}
\nonumber
\Upsilon_{1}&=\sum_{\theta_{1}=0}^{\infty}{\sum_{\theta_{3}=0}^{P-1}}{\sum_{{\theta_{4}}=0}^{N{(\theta_{3}+1)}-1}}\sum_{s_{1}=0}^{\infty}{\sum_{\tau_{\theta_{4}}}}\frac{\lambda_{{11},\theta_{{1},{3},{4}}}(-\lambda_{14})^{s_{1}}}{ln(2)s_{1}!}
\int_{0}^{\infty}G_{2,2}^{1,2}\left[\xi_{db}\Biggl|\begin{array}{c}
1,1\\
1,0\\
\end{array}
\right]
G_{0,1}^{1,0}\left[\lambda_{15}{\xi_{db}}^{\frac{\alpha_{b}}{2}}\Biggl|\begin{array}{c}
-\\
0\\
\end{array}
\right]\xi_{db}^{s_1}
\\\label{54}
&\times{{\Biggl({\sum_{\theta_{7}=0}^{\theta_{1}}{\lambda_{{12},\theta_{7}}}\xi_{db}^{r_1}+{\sum_{\theta_{8}=0}^{\mu_{b}-1}{\lambda_{{13},\theta_{8}}}\xi_{db}^{r_2}}\Biggl)}}}d{\xi_{db}}.
\end{align}
Now, integrating \eqref{54} using \cite[eq.~2.24.1.1,]{prudnikov1988integrals}, we get 
\begin{align}
\nonumber
\Upsilon_{1}&=\sum_{\theta_{1}=0}^{\infty}{\sum_{\theta_{3}=0}^{P-1}}{\sum_{{\theta_{4}}=0}^{N{(\theta_{3}+1)}-1}}\sum_{s_{1}=0}^{\infty}\sum_{\theta_{7}=0}^{\theta_{1}}{\sum_{\tau_{\theta_{4}}}}z_{1}G_{ 2\alpha_{b}, 2(1+\alpha_{b})}^{2(1+\alpha_{b}), \alpha_{b}}
\left[\left(\frac{\lambda_{15}}{2}\right)^{2}\Biggl|\begin{array}{l}
\Delta(\alpha_{b},-{c_{1}}-1),\Delta(\alpha_{b},-{c_{1}})  \\
\Delta(2,0), \Delta(\alpha_{b},-{c_{1}}-1),\Delta(\alpha_{b},-{c_{1}}-1)
\end{array}
\right]
\\
\label{55}
&+\sum_{\theta_{1}=0}^{\infty}{\sum_{\theta_{3}=0}^{P-1}}{\sum_{{\theta_{4}}=0}^{N{(\theta_{3}+1)}-1}}\sum_{s_{1}=0}^{\infty}\sum_{\theta_{8}=0}^{\mu_{b}-1}{\sum_{\tau_{\theta_{4}}}}z_{2}G_{ 2\alpha_{b}, 2(1+\alpha_{b})}^{2(1+\alpha_{b}), \alpha_{b}}
\left[\left(\frac{\lambda_{15}}{2}\right)^{2}\Biggl|\begin{array}{l}
\Delta(\alpha_{b},-{c_{2}}-1),\Delta(\alpha_{b},-{c_{2}})  \\
\Delta(2,0), \Delta(\alpha_{b},-{c_{2}}-1),\Delta(\alpha_{b},-{c_{2}}-1)
\end{array}
\right],
\end{align}
where $z_1=\frac{\sqrt{2}{\lambda_{{12},\theta_{7}}}\lambda_{{11},\theta_{{1},{3},{4}}}(-\lambda_{14})^{s_{1}}}{\alpha_{b}(2\pi)^{{\alpha_b}-\frac{1}{2}}ln(2)s_{1}!}$, ${c_{1}}={r_{1}}+{s_{1}}$, ${c_{2}}={r_{2}}+{s_{1}}$ and $z_2=\frac{\sqrt{2}{\lambda_{{13},\theta_{8}}}\lambda_{{11},\theta_{{1},{3},{4}}}(-\lambda_{14})^{s_{1}}}{\alpha_{b}(2\pi)^{{\alpha_b}-\frac{1}{2}}ln(2)s_{1}!}$.
\subsection{Derivation of $\Upsilon_{2}$}
The similar process of deriving $\Upsilon_{1}$ can be followed to obtain the expression of $\Upsilon_{2}$, which is given as follows:
\begin{align}
\nonumber
\Upsilon_{2}&=\sum_{\theta_{2}=0}^{\infty}{\sum_{\theta_{11}=0}^{QN-1}}{\sum_{{\theta_{12}}=0}^{\theta_{2}}}\sum_{s_{2}=0}^{\infty}{\sum_{\tau_{\theta_{11}}}}z_{3}G_{ 2\alpha_{e}, 2(1+\alpha_{e})}^{2(1+\alpha_{e}), \alpha_{e}}\Biggl[\left(\frac{\hbar_{14}}{2}\right)^{2}
\Biggl|\begin{array}{l}
\Delta(\alpha_{e},-{c_{3}}-1),\Delta(\alpha_{e},-{c_{3}})  \\
\Delta(2,0), \Delta(\alpha_{e},-{c_{3}}-1),\Delta(\alpha_{e},-{c_{3}}-1)
\end{array}
\Biggl]
\\
\label{56}
&+\sum_{\theta_{2}=0}^{\infty}{\sum_{\theta_{11}=0}^{QN-1}}{\sum_{{\theta_{12}}=0}^{\mu_{e}-1}}\sum_{s_{2}=0}^{\infty}{\sum_{\tau_{\theta_{11}}}}z_{4}G_{ 2\alpha_{e}, 2(1+\alpha_{e})}^{2(1+\alpha_{e}), \alpha_{e}}\Biggl[\left(\frac{\hbar_{14}}{2}\right)^{2}
\Biggl|\begin{array}{l}
\Delta(\alpha_{e},-{c_{4}}-1),\Delta(\alpha_{e},-{c_{4}})  \\
\Delta(2,0), \Delta(\alpha_{e},-{c_{4}}-1),\Delta(\alpha_{e},-{c_{4}}-1)
\end{array}
\Biggl],
\end{align}
where $z_3=\frac{\sqrt{2}{\hbar_{{9},\theta_{12}}}\hbar_{{12},\theta_{{2},{11}}}(-\hbar_{13})^{s_{2}}}{\alpha_{e}(2\pi)^{{\alpha_e}-\frac{1}{2}}ln(2)s_{2}!}$, ${c_{3}}={r_{3}}+{s_{2}}$, ${c_{4}}={r_{4}}+{s_{2}}$ and $z_4=\frac{\sqrt{2}{\hbar_{{10},\theta_{13}}}\hbar_{{12},\theta_{{2},{11}}}(-\hbar_{13})^{s_{2}}}{\alpha_{e}(2\pi)^{{\alpha_e}-\frac{1}{2}}ln(2)s_{2}!}$. Finally substituting the values of \eqref{55} and \eqref{56} into \eqref{50}, we get the analytical expression of ESMC.

\subsection{Generalization offered by the ESMC expression}
Due to the generic nature of $\alpha-\mu$ fading channel, the derived expression of ESMC of the dual-hop multicast model is generalized too. This helps to obtain more useful insights, since analyzing secrecy performance of the proposed model unifies the performance evaluation of a wide range of existing multipath models. Such as utilizing ESMC expression, we can easily reproduce the results of \cite[eq.~25,]{huang2017secrecy} for $N=1$, $P=1$, $\alpha=2$ and $\mu=1$, and \cite[eq.~31,]{bankey2019physical} for $N=1$, $P=1$, $\alpha=2$, and $\mu=m$.

\section{PNSMC Analysis}
\label{sec5}
The PNSMC is a significant benchmark for the demonstration of physical layer security performance. This metric indicates that secure communication between the source and legitimate receiver is possible only if the quality of the legitimate link is better than the illegitimate link i.e. the secrecy capacity is a positive quantity. PNSMC is defined as \cite[eq.~25]{islam2020secrecy}
\begin{align}
\label{pnsmc1}
P(\mathcal{C}_{s}>0)&=\int_{0}^{\infty}\left(\int_{0}^{\xi_{db}}f_{\zeta_{max}}(\xi_{de})d{\xi_{de}}\right) f_{\zeta_{min}}(\xi_{db}) d{\xi_{db}}.
\end{align}
Substituting \eqref{35} and \eqref{41} into \eqref{pnsmc1}, and then performing integration, we can derive the closed-form expression of PNSMC following the similar procedures of obtaining SOPM expression. But this derivation will not exhibit a significant contribution since the PNSMC can be easily calculated from the SOPM by considering a zero target rate as the following \cite[eq.~34,]{badrudduza2020enhancing}:
\begin{align}
\label{pnsmc2}
  P(\mathcal{C}_{s}>0)=1-P_{out}(\varphi_{c})\bigg|_{\varphi_{c}=0} .
\end{align}
Hence, we set $\varphi_{c}=0$ in \eqref{sop} and then substitute this value in \eqref{pnsmc2} to get the values of PNSMC.

\section{Numerical Outcomes}
\label{sec6}

The analytical expressions of SOPM, ESMC, and PNSMC are numerically illustrated in this section in terms of the environmental fading and shadowing condition with multiple relays, users, and eavesdroppers. Although, the final closed form expressions in \eqref{sop}, \eqref{55}, and \eqref{56} contain infinite sums, the values of the expressions converges accurately for first 20 terms. The accuracy of the expressions of the performance metrics mentioned above is validated through Monte-Carlo simulation via MATLAB which shows a perfect match with the analytical results.

\begin{figure}[!ht]
\vspace{-20mm}
    \centerline{\includegraphics[width=0.7\textwidth]{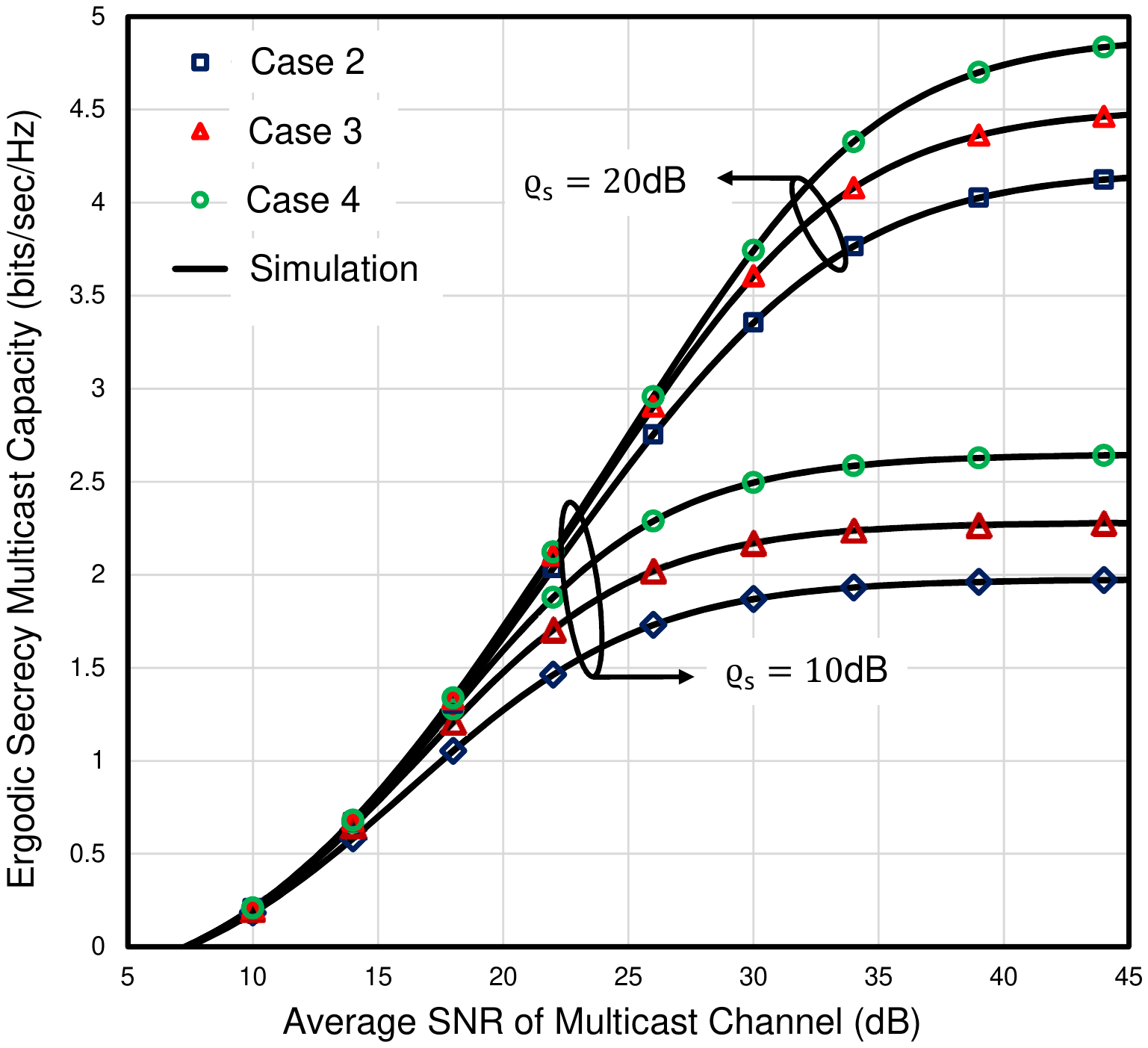}}
    \vspace{-72mm}
    \caption{ESMC versus $\varrho_{b}$ for selected values of $\varrho_{s}$ with $\alpha_{b}= 2$, $\alpha_{e}=2$, $\mu_{b}=1$, $\mu_{e}=1$, $N=5$, $P=5$, $Q=5$, and $\varrho_{e}=-10$ dB.}
    \label{fig:2}
\end{figure}
The effect of shadowing on the ESMC is presented in Fig. \ref{fig:2} considering two different channel conditions in the $\mathcal{S}-\mathcal{N}$ link. Here, case 2 represents severe shadowing whereas case 4 illustrates the minimum amount of shadowing. It is clearly observed that the ESMC of the system decreases rapidly with the shadowing as the shadowing tends to hinder the proper communication between the source and legitimate receivers. In that case, the secrecy capacity can still be improved easily by increasing the average SNR of the $\mathcal{S}-\mathcal{N}$ link as testified in \cite{bankey2019physical,huang2017secrecy}. The simulated results also show a tight match with the analytical results which ensure the validity of the derived expressions.

\begin{figure}[!ht]
\vspace{-20mm}
    \centerline{\includegraphics[width=0.7\textwidth]{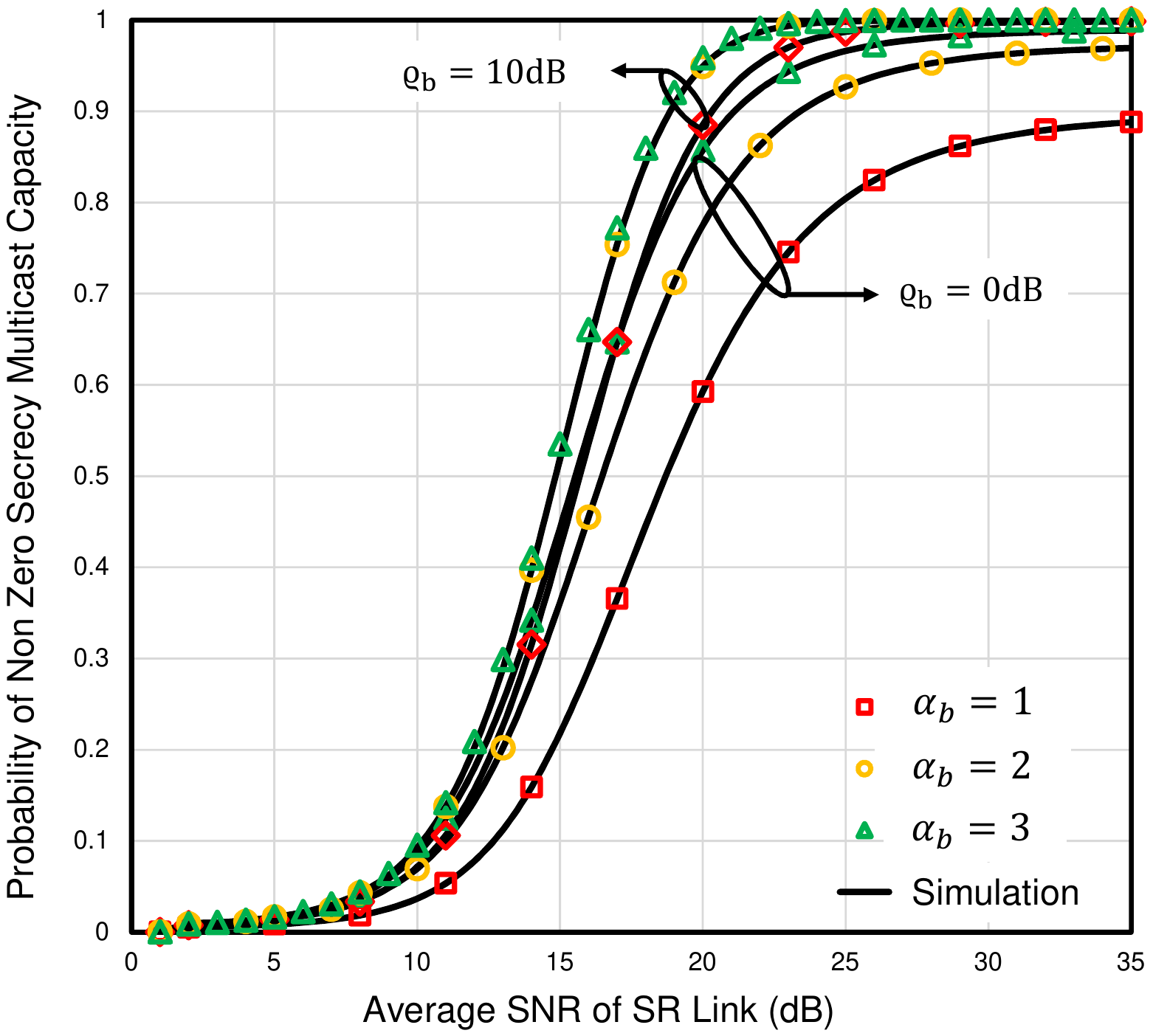}}
    \vspace{-72mm}
    \caption{PNSMC versus $\varrho_{s}$ considering Case-1 for selected values of $\alpha_{b}$ and $\varrho_{b}$ with  $\alpha_{e}=2$, $\mu_{b}=1$, $\mu_{e}=1$, $N=5$, $P=5$, $Q=5$, and $\varrho_{e}=-10$ dB.}
    \label{fig:4}
\end{figure}

\begin{figure}[!ht]
\vspace{-20mm}
    \centerline{\includegraphics[width=0.7\textwidth]{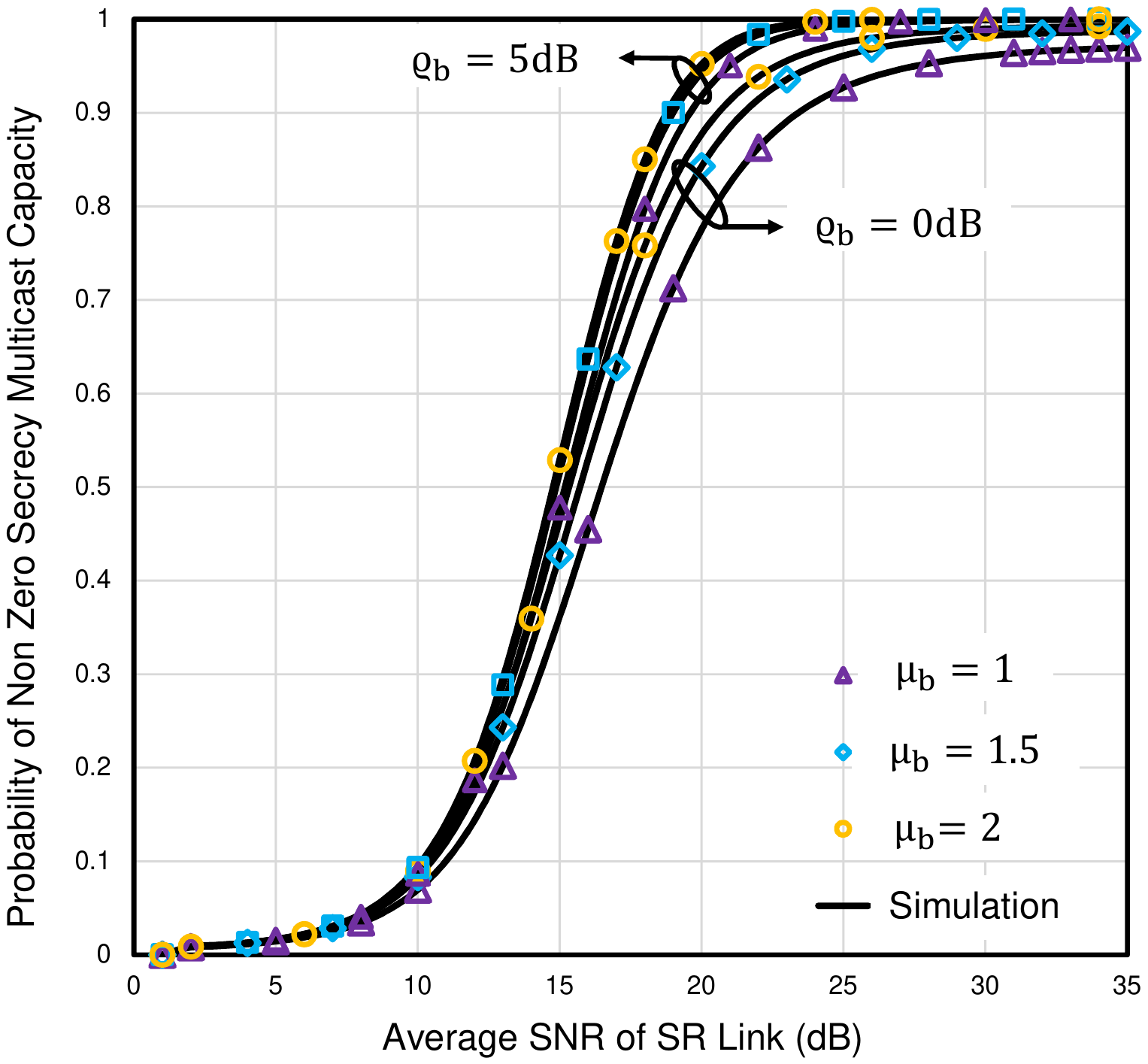}}
    \vspace{-75mm}
    \caption{PNSMC versus $\varrho_{s}$ for selected values of $\mu_{b}$ and $\varrho_{b}$ with Case-1 considering $\alpha_{b}= 2$, $\alpha_{e}=2$, $\mu_{e}=1$, $N=5$, $P=5$, $Q=5$,  and $\varrho_{e}=-10$ dB.}
    \label{fig:5}
\end{figure}

The PNSMC of the proposed system is presented in Figs. \ref{fig:4} and \ref{fig:5} with respect to (w.r.t.) $\varrho_{s}$ where the effect of fading severity of the $\mathcal{N}-\mathcal{P}$ link is thoroughly demonstrated. It is observed that the security of the system can be enhanced by uplifting the average SNR at the second hop of the proposed model. One of the most common causes of the degradation in PNSMC is the fading condition of the communication channel. But the effect of fading can be overcome by increasing the non-linearity and clustering parameters ($\alpha_{b}$, $\mu_b$) which actually decreases the total environmental fading of the multicast channels. Due to this reason, the receivers can easily receive the information sent from the relay,  and accordingly, the PNSMC increases. Similar performance was also seen in \cite{kong2018secrecy, badrudduza2021security} which substantially verify our outcomes.

\begin{figure}[!ht]
\vspace{-20mm}
    \centerline{\includegraphics[width=0.7\textwidth]{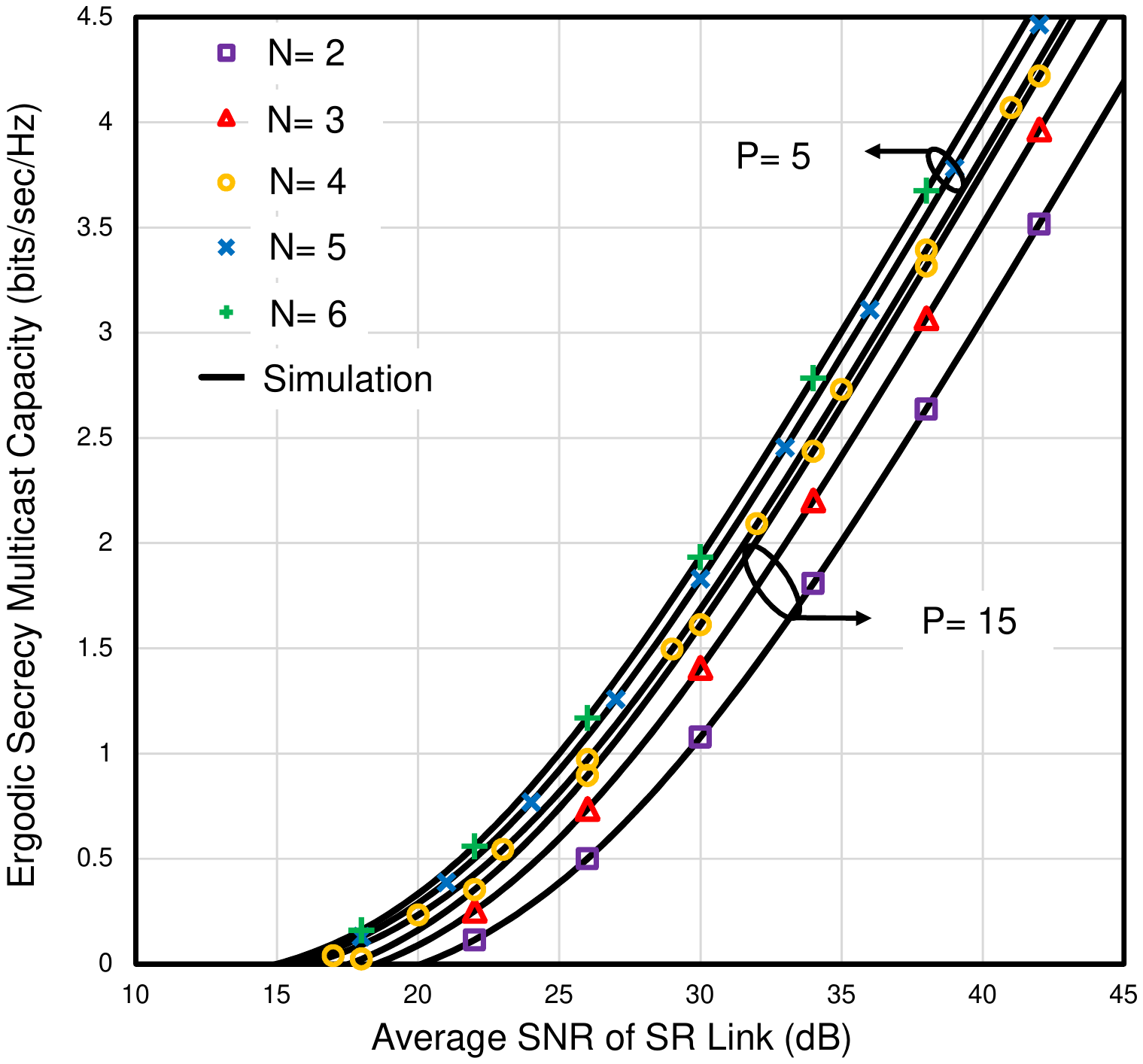}}
    \vspace{-72mm}
    \caption{ESMC versus $\varrho_{b}$ illustrating the effect of $N$ and $P$ for Case-1 with $\alpha_{b}= 2$, $\alpha_{e}=2$, $\mu_{b}=1$, $\mu_{e}=1$, $N=$, $P=$, $Q=5$, and $\varrho_{e}=-10$ dB.}
    \label{fig:3}
\end{figure}
\begin{figure}[!ht]
\vspace{-20mm}
    \centerline{\includegraphics[width=0.7\textwidth]{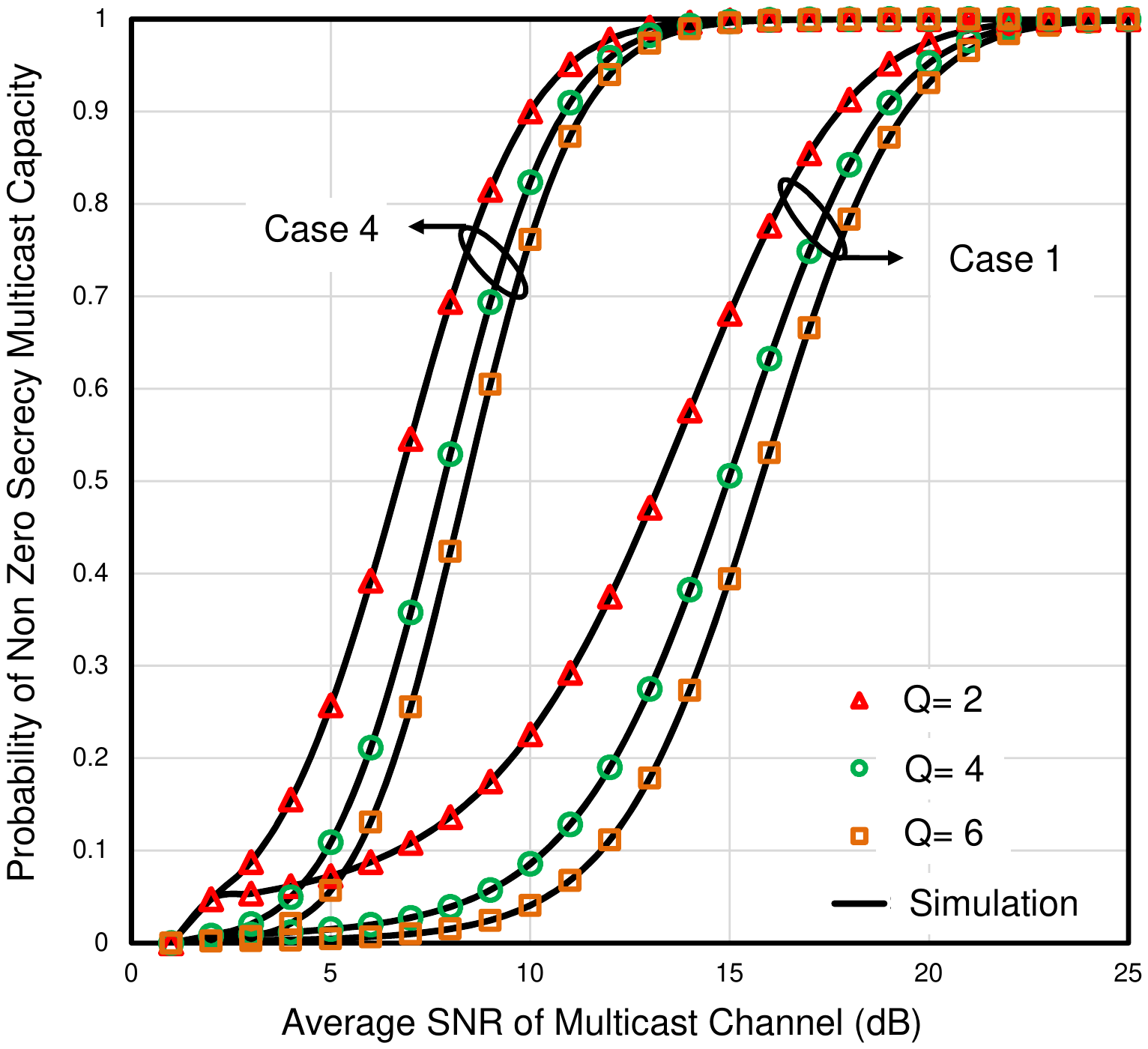}}
    \vspace{-72mm}
    \caption{PNSMC versus $\varrho_{b}$ for different values of $Q$ and shadowing conditions with $\alpha_{b}=2 $, $\alpha_{e}=2$, $\mu_{b}=1$, $\mu_{e}=1$, $N=5$, $P=5$, and $\varrho_{e}=-10$ dB.}
    \label{fig:6}
\end{figure}
We know that with the increasing number of receivers, the bandwidth for each receiver reduces because the total bandwidth is fixed. As a result, the ESMC of the system decreases sharply with the number of receivers. This phenomenon is graphically represented in Fig. \ref{fig:3} and it can be easily seen that the ESMC is lower for the larger number of receivers. A solution to this problem is also illustrated in this figure in the form of number of relays. Similar to \cite{bao2013cognitive,chen2011exact}, the increment in the number of relays also upgrades the ESMC of the system because a higher number of relays increases the probability of transmitting a better signal from the best relay.

In Fig. \ref{fig:6}, the PNSMC is plotted w.r.t. $\varrho_{s}$ to show the unfavorable effects of shadowing and increasing number of eavesdroppers on the system's security. As severe shadowing inhibits proper information transfer, PNSMC will decrease. If a large number of eavesdroppers are also present in the system, they will be able to steal more data and make the system vulnerable and hence the PNSMC will decrease regardless of the shadowing condition. This result also matches well with the corresponding results of \cite{badrudduza2020enhancing}.

\begin{figure}[!ht]
\vspace{-15mm}
    \centerline{\includegraphics[width=0.7\textwidth]{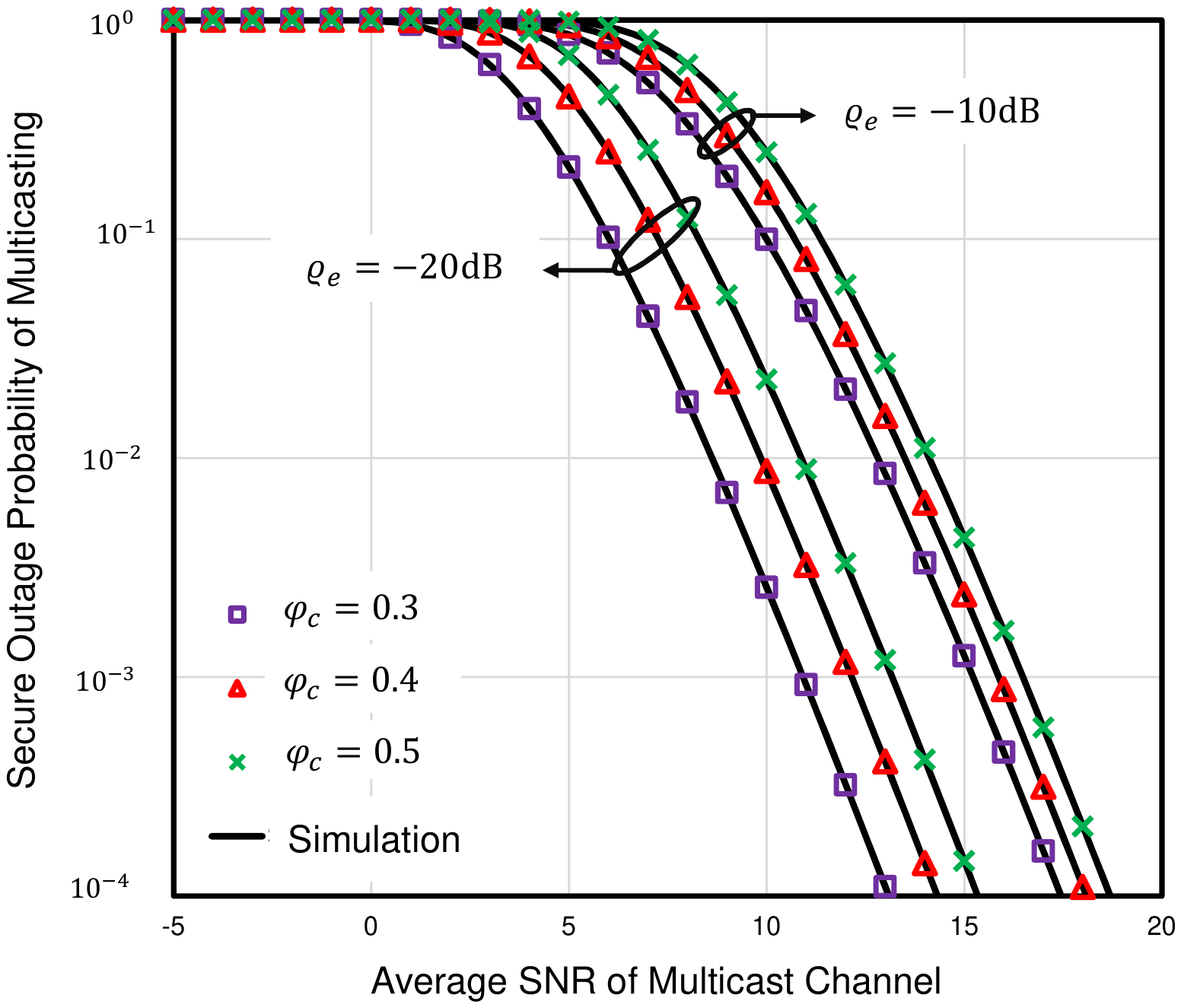}}
    \vspace{-80mm}
    \caption{SOPM versus $\varrho_{b}$ with different values of $\varphi_{c}$ and $\varrho_{e}$ with $\alpha_{b}= 2$, $\alpha_{e}=2$, $\mu_{b}=1$, $\mu_{e}=1$, $N=5$, $P=5$, and $Q=5$.}
    \label{fig:7}
\end{figure}
\begin{figure}[!ht]
\vspace{-20mm}
    \centerline{\includegraphics[width=0.7\textwidth]{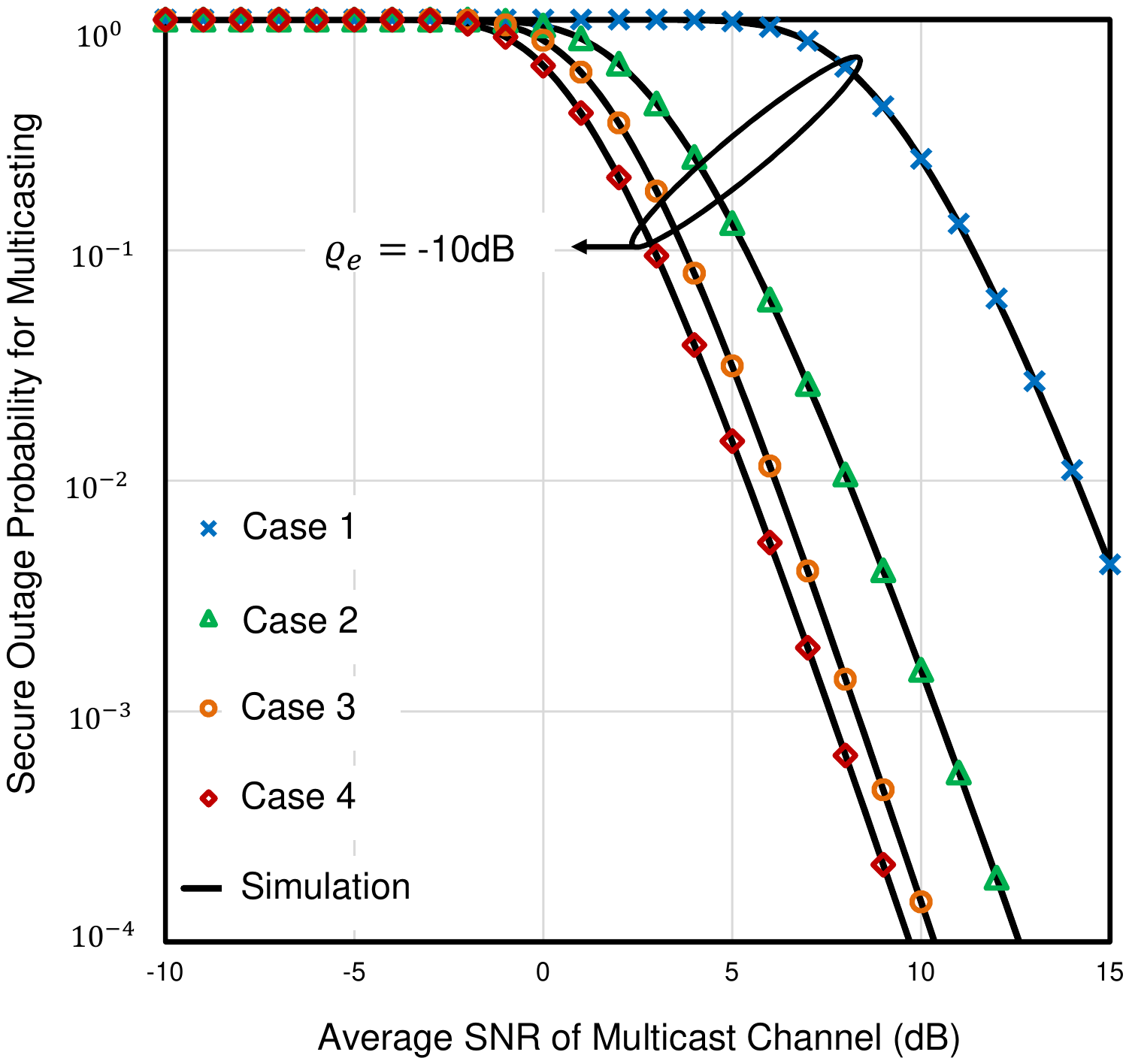}}
    \vspace{-80mm}
    \caption{SOPM versus $\varrho_{b}$ for various shadowing conditions from Table \ref{table1} with $\alpha_{b}= 2$, $\alpha_{e}=2$, $\mu_{b}=1$, $\mu_{e}=1$, $N=5$, $P=5$, $Q=5$, $\varphi_{c}=0.5$, and $\varrho_{e}=-10$ dB.}
    \label{fig:8}
\end{figure}
Figure \ref{fig:7} represents the SOPM for the proposed system which is plotted against the $\varrho_{b}$. The effect of $\varrho_{e}$ and $\varphi_{c}$ on the secrecy performance of the system is illustrated in this figure. It is clearly seen that the increment in both $\varrho_{e}$ and $\varphi_{c}$ has a negative impact on the system's security. Because the increment in $\varrho_{e}$ decreases the fading of the eavesdropper link which helps the eavesdroppers to steal a large amount of data as affirmed in \cite{moualeu2019transmit,bankey2017secrecy}. On the other hand, the raise in $\varphi_{c}$ requires a higher transmission rate which increases the probability of outage of the system \cite{9354969}.

The effect of four shadowing conditions (Table \ref{table1}) on SOPM is illustrated in Fig. \ref{fig:8} w.r.t. $\varrho_{b}$. From this figure, it is comprehensible that for deep shadowing (Case 1), SOPM is higher with constant values of $\varrho_{e}$ (=-10 dB). But in case 4, SOPM is lower as light shadowing is less detrimental for secure data transfer. Similar conclusions on the impact of shadowing over the SR model were also drawn in \cite{guo2016secure,bankey2019physical} which clearly authenticates our results.

\textbf{Generic Property of the Proposed Model}:
\begin{figure}[!ht]
\vspace{-20mm}
    \centerline{\includegraphics[width=0.7\textwidth]{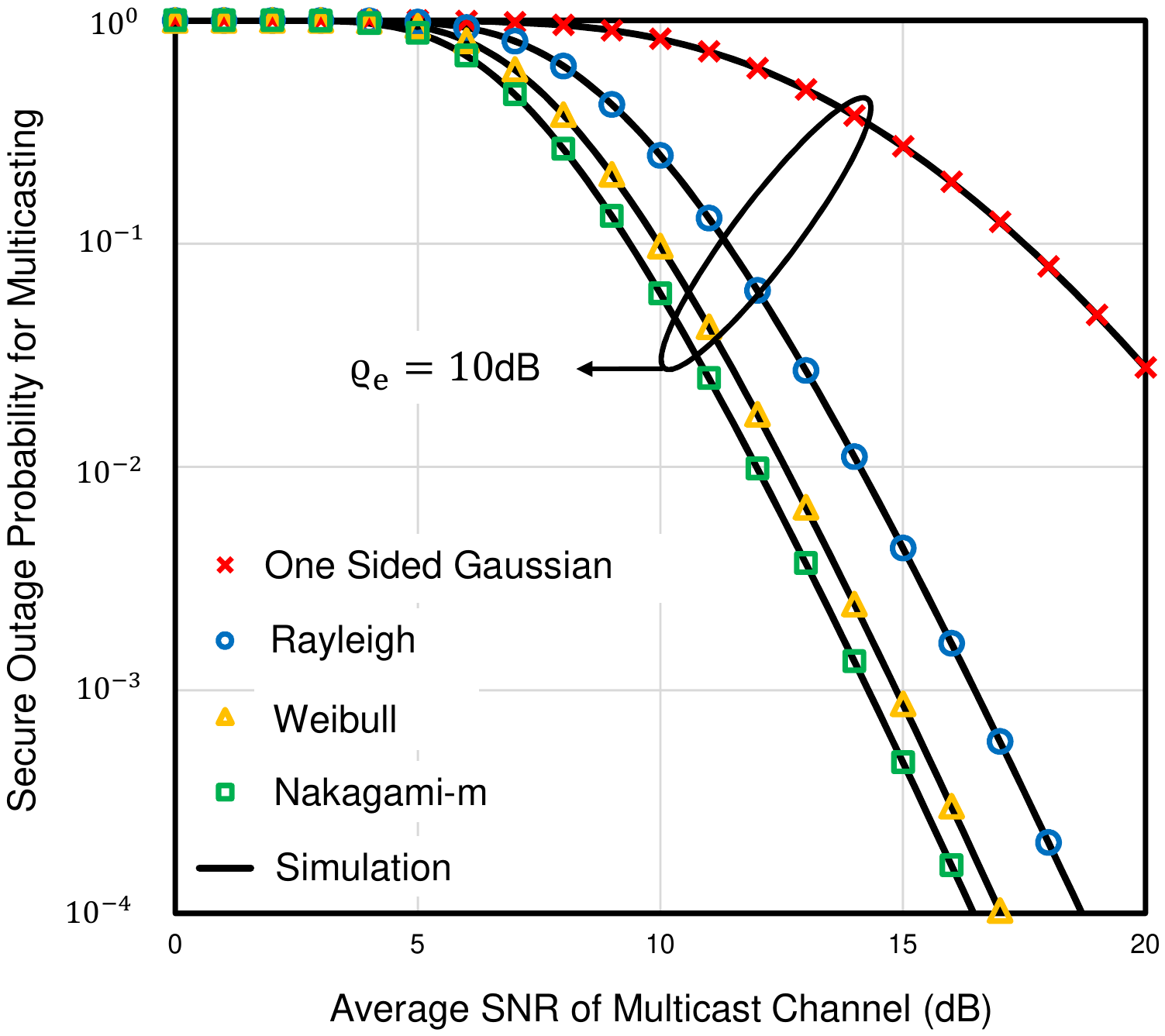}}
    \vspace{-80mm}
    \caption{Generic property of the proposed model for Case-1 with $N=5$, $P=5$, $Q=5$, and $\varphi_{c}=0.5$.}
    \label{fig:9}
\end{figure}
The generalized fading channels has an inherent advantage over the multipath channels that it can naturally represent multiple multipath fading channels just by varying the value of some physical parameters. As this proposed work considers generalized $\alpha-\mu$ fading channel at the second hop, secrecy performance of multiple HSTRN can be evaluated from this model which is demonstrated in Fig. \ref{fig:9}. It is noteworthy that the secrecy performance of the SR-Nakagami-$m$ fading model was evaluated in \cite{bankey2019physical,bankey2017secrecy}which can be replicated by our proposed model for $\alpha=2$ and $\mu=m$. A similar analysis was again done in \cite{cao2018relay,huang2017secrecy} for the SR-Rayleigh fading model which can also be easily represented by this proposed work considering $\alpha=2$ and $\mu=1$. Further, the security analysis of SR-Weibull ($\alpha=3$, $\mu=1$) fading model and SR-One sided Gaussian ($\alpha=1$, $\mu=1$) model can be examined by this proposed work which is yet to be reported in any research. Thus, this proposed model shows complete novelty compared to the existing works.

\section{Conclusion}
\label{sec7}
This paper deals with the secrecy performance analysis and mathematical modeling of a dual-hop HSTRN with SR and $\alpha$-$\mu$ fading channels at first and second hops, respectively. Mathematically tractable analytical expressions of SOPM, ESMC, and PNSMC are acquired to examine the effect of shadowing on the system's secrecy performance. We further justify those expressions via computer simulations. Although shadowing has a negative impact on the system’s secrecy performance, the numerical results show that the opportunistic relaying scheme can be a remedy to this situation. Besides, this scheme is also equally  useful in overcoming the detrimental impacts of fading, multicast users, and multiple eavesdroppers. The derived generalized expressions can also be applied to analyze the security performance of some well-known hybrid networks, e.g. SR-Nakagami-$m$, SR-Weibull, SR-Rayleigh, and SR-one-sided Gaussian as special cases which demonstrate an enormous versatility of the proposed scheme over existing works.


\bibliography{sample}

\end{document}